\documentclass[amsmath,amssymb,aps]{revtex4}

\selectfont 

\usepackage{graphicx}       
\usepackage{dcolumn}        
\usepackage{bm}             
\usepackage{color}
\usepackage{multirow}

\usepackage{float} 

\newcommand{\beq}{\begin{equation}}
\newcommand{\eeq}{\end{equation}}
\newcommand{\beqy}{\begin{eqnarray}}
\newcommand{\eeqy}{\end{eqnarray}}

\def\be{\begin{equation}}
\def\ee{\end{equation}}
\def\bea{\begin{eqnarray}}
\def\eea{\end{eqnarray}}

\begin{document}

\title{Quasinormal modes of black strings in de Rham-Gabadadze-Tolley massive gravity}

\author{Supakchai Ponglertsakul}
\email{supakchai.p@gmail.com}
\affiliation{High Energy Physics Theory Group, Department of Physics, Faculty of Science,
Chulalongkorn University, Phyathai Rd., Bangkok 10330, Thailand}

\author{Piyabut Burikham}
\email{piyabut@gmail.com}
\affiliation{High Energy Physics Theory Group, Department of Physics, Faculty of Science,
Chulalongkorn University, Phyathai Rd., Bangkok 10330, Thailand}

\author{Lunchakorn Tannukij}
\email{l_tannukij@hotmail.com}
\affiliation{Department of Physics, Hanyang University, Seoul 133-891, South Korea}

\date{\today}

\begin{abstract}
The effect of massive scalar perturbations on neutral black string in de Rham-Gabadadze-Tolley (dRGT) massive gravity is investigated through the study of the quasi-normal modes~(QNMs).  Due to the similarity between the equation of motion of the field in the black-string and black-hole background, similar numerical and analytical techniques can be used to explore the behaviour of the QNMs.  We use the asymptotic iteration method (AIM) and the WKB method to numerically calculate the QNMs of scalar perturbation in the black string background with positive cosmological constant. High-momentum behaviour of such QNMs can be analytically approximated by the first-order WKB method with excellent accuracy.  For near-extremal black string with event horizon very close to the cosmological horizon, the P\"oschl-Teller technique gives accurate analytic formula for the QNMs. When massive-gravity-parameter $\gamma$ increases, we found that the scalar modes oscillate with higher frequencies and decay faster. The QNMs of black string in spacetime with negative cosmological constant are explored in all range of possible $\gamma$ using the spectral method.  We found the movement of the holographic sound poles to collide and form diffusive poles as $\gamma$ changes from positive to negative values. We observe no evidence of instability of neutral black string in both positive and negative cosmological constant cases.

\end{abstract}

\pacs{nnnnn}

\maketitle

\section{Introduction}\label{sec:intro}

A possibility of whether graviton can be massive has been a question for a long time in physics community since the birth of the theory of propagating massless graviton known as the Einstein's general relativity. There were many attempts to generalize the Einstein's theory to describe propagating massive graviton. In 1939, Fierz and Pauli first successfully constructed a linear theory of massive gravity \cite{Fierz:1939ix} by adding a suitable mass term parameterized by graviton mass to the linearized version of general relativity. Though adding the mass term seems to break the general covariance, it can be recovered via the use of the so-called St\"uckelberg trick (see \cite{Hinterbichler:2011tt} for a review on how to apply the trick to the massive gravity model). However, thanks to the nonvanishing mass, the linear theories of massive gravity suffer from the so-called van Dam-Veltman-Zakharov discontinuity \cite{vanDam:1970vg,Zakharov:1970cc} which corresponds to the fact that predictions from the massive gravity in the massless limit do not agree with the predictions from the general relativity. Even though a hope for fixing such discontinuity problems lies in a nonlinear regime of massive gravity \cite{Vainshtein:1972sx}, generic nonlinear generalization from the Fierz-Pauli linear massive gravity usually possesses the so-called Boulware-Deser ghost mode in the theory \cite{Boulware:1973my}. Fortunately, a special class of nonlinear theory of massive gravity has been successfully found by de Rham, Gabadadze, and Tolley, dubbed as dRGT massive gravity theory \cite{deRham:2010ik,deRham:2010kj}. The dRGT massive gravity has proved itself useful in cosmology since its cosmological solution incorporates the graviton mass as an effective cosmological constant \cite{Gumrukcuoglu:2011ew,Gumrukcuoglu:2011zh}, hence the accelerating expansion of the universe can be explained in this context.

Four dimensional static spherically symmetric black holes in dRGT massive gravity with and without $U(1)$ charge are constructed explicitly in \cite{Ghosh:2015cva}. The effective cosmological constant emerges naturally as a result of graviton mass term. Therefore, these black holes can be considered as generalized Schwarzschild/Reissner-Nordstr\"om black holes with positive and negative cosmological constant. The temperature, entropy and thermodynamic stability of these dRGT black holes are also explored \cite{Ghosh:2015cva}. Stability of the Schwarzschild-de Sitter black hole in dRGT massive gravity is studied in Ref.~\cite{Kodama:2013rea}.  Stability of black holes in bi-gravity extension of the dRGT model is studied for certain specific conditions in Ref.~\cite{Babichev:2013una,Babichev:2014oua}.  Beyond spherical symmetry, a cylindrically symmetric solution with an event horizon is possible even in standard general relativity. The cylindrical black hole (or black string) is proposed by Lemos with the presence of negative cosmological constant \cite{Lemos:1994xp}. In higher dimensions, black branes and strings can exist and their stability can be explored using the QNMs~\cite{Konoplya:2008yy}.  As a generalization of Lemos's black string, 
static solution with cylindrical symmetry is found in dRGT massive gravity \cite{Tannukij:2017jtn}. Unlike the black string in general relativity, there exists dRGT black string solution even with a positive cosmological constant effectively contributed by the graviton mass. 

On the other hand, linear stability of black hole closely relates to a study of field perturbation exterior to the black hole's horizon. The evolution of field perturbation is dictated by gravitational interaction between black hole and the test-field itself. The existence of event horizon requires that the field mode near the black hole must satisfy with ingoing wave condition. Thus, the perturbation modes are not normal modes but rather quasinormal modes (QNMs). The quasinormal spectrum will be complex frequencies that can be uniquely determined by the black hole's mass, charge and angular momentum. For instance, the QNMs of neutral and charged black holes in dRGT massive gravity are computed \cite{Prasia:2016fcc}. In our previous work, we investigate a rich structure of dRGT massive gravity by calculating the effect of massive charged scalar perturbation on charged dRGT black hole \cite{Burikham:2017gdm}. While stable modes are discovered in some parameter space, we also find that some dRGT black holes suffer from superradiant instability \cite{Burikham:2017gdm}. Interestingly, the QNMs can be categorized into three families, $a)$ the near event horizon mode, $b)$ the near cosmic horizon mode, $c)$ all-region mode \cite{Burikham:2017gdm}.  Similar behaviours are found for the QNMs of black string in the dRGT background, the near-horizon modes of the event horizon and cosmic horizon for positive cosmological constant case are purely diffusive~(or negative imaginary) but we will present more detailed analysis elsewhere.  In this article, we will focus on the all-region QNMs of the black string.  We refer interested readers to Refs. \cite{Kokkotas:1999bd,Berti:2009kk,Konoplya:2011qq} for very nice reviews on black hole QNMs.

Black hole, black brane and black string in asymptotically anti de-Sitter (AdS) space with negative cosmological constant are also interesting from the holographic duality viewpoint.  They are dual to the thermal field theory on the boundary whose temperature is identified with the Hawking temperature of the spacetime.  Existence of horizon implies finite temperature of the dual field theory and simultaneously implies that any fluctuations in such background would inevitably produce disssipations into the horizon.  Such dissipations are encoded in the QNMs of the fluctuations of the spacetime.  Due to the holographic duality, these QNMs are exactly the poles of the retarded thermal Green function of the dual field theory.  They represent the relaxation time of the perturbed thermal field system in the dual picture to reach the thermal equilibrium again.  Massive graviton in the bulk induces the breaking of diffeomorphism-invariance on the boundary and originates the momentum dissipation in the dual hydrodynamics~\cite{Davison:2013jba,Davison:2014lua}.  

For cylindrically symmetric set-up, toroidal, cylindrical and planar black holes in general relativity are stable against small perturbation in all channels, scalar, vector and tensor \cite{Cardoso:2001vs}. However, the stability of black string in the dRGT massive gravity has never been addressed. Therefore in this paper, we attempt to investigate this problem systematically. In Sec.~\ref{sec:formal}, the basic set-up for constructing neutral black string in dRGT massive gravity is introduced. Then, we explore the effects of linear term $(\gamma)$ on the horizon structure of black strings in both positive and negative cosmological constant scenario. In Sec.~\ref{sec:pert}, we examine massive scalar perturbation on the dRGT black string background. The Klein-Gordon equation on curve spacetime and boundary conditions are discussed. It is possible to obtain semi-analytical formula for the quasinormal frequency in a small-universe near-extremal limit, this is explored in Sec.~\ref{sec:nearEx}. The QNMs of dRGT black string with positive cosmological constant are calculated and discussed in Sec.~\ref{sec:dS}. High-momentum behaviour of QNMs with positive cosmological constant is numerically studied in Sec. \ref{numre} and analytically approximated using the first-order WKB method in Sec. \ref{wkbhk}.  In Sec.~\ref{sec:ads}, we present the QNMs of dRGT black string with negative cosmological constant. Finally we summarize our results in Sec.~\ref{sec:conclusion}.

\section{Formalism}\label{sec:formal}

We shall consider neutral black string in dRGT massive gravity couples with massive scalar field. This is described by the action \cite{deRham:2010kj} (with $c=8\pi G=1$)
\beq
S = \frac{1}{2}\int d^4 x \sqrt{-g}\left[R + m^2_g \mathcal{U}(g,\phi^a)- g^{\mu\nu}\nabla_{(\mu} \Phi \nabla_{\nu)} \Phi - m_s^2\Phi^2 \right],
\eeq
where $m_g$ and $m_s$ can be considered as the graviton mass and the scalar field mass. The ghost-free massive graviton self-interacting potential is defined by
\beq
\mathcal{U}(g,\phi^a) = \mathcal{U}_2 + \alpha_3 \mathcal{U}_3 + \alpha_4 \mathcal{U}_4
\eeq
where
\begin{align}
\alpha_3 &= \frac{\alpha-1}{3}, \\
\alpha_4 &= \frac{\beta}{4} + \frac{1-\alpha}{12}. \\
\mathcal{U}_2 &= [\mathcal{K}]^2 - [\mathcal{K}^2], \\
\mathcal{U}_3 &= [\mathcal{K}]^3 - 3[\mathcal{K}][\mathcal{K}^2] + 2[\mathcal{K}^3], \\
\mathcal{U}_4 &= [\mathcal{K}]^4 - 6[\mathcal{K}]^2[\mathcal{K}^2] + 8[\mathcal{K}][\mathcal{K}^3] + 3[\mathcal{K}^2]^2 - 6[\mathcal{K}^4].
\end{align}
$\alpha$ and $\beta$ are free parameters. $K^{\mu}_{\nu} = \delta^{\mu}_{\nu} - \sqrt{g^{\mu\sigma}f_{ab}\partial_{\sigma}\phi^a\partial_{\nu}\phi^b}$. $[\mathcal{K}]=\mathcal{K}^{\mu}_{\mu}$ and $[\mathcal{K}^n]=(\mathcal{K}^n)^{\mu}_{\mu}$. We will work in unitary gauge for which the four St\"{u}ckelberg fields take the form $\phi^a = x^{\mu}\delta^a_{\mu}$. The fiducial metric $f_{ab}$ is chosen to be 
\beq
f_{ab} = diag(0,0,\alpha_g^2 h_0^2,h_0^2),  \label{fideq}
\eeq
for the coordinates $(t,r,z,\varphi)$ where $\alpha_g$ and $h_0$ are arbitrary constants. The parentheses denote symmetrization $X_{(ab)} = \frac{1}{2} \left(X_{ab} + X_{ba}\right)$. 

\subsection{Field equations}
\textbf{1. Einstein's equations:}
\beq
R_{\mu\nu} - \frac{1}{2} R g_{\mu\nu} = -m^2_g X_{\mu\nu} + T_{\mu\nu}^\Phi  \label{einstein-eq}
\eeq
where the full expression for $X_{\mu\nu}$ can be found in \cite{Ghosh:2015cva,Tannukij:2017jtn}. The energy-momentum tensor is given by
\begin{eqnarray}
T_{\mu\nu}^{\Phi} &=& \nabla_{(\mu} \Phi \nabla_{\nu)} \Phi - g_{\mu\nu}\left(g^{\sigma\rho}\nabla_{(\sigma} \Phi \nabla_{\rho)} \Phi + m_s^2\Phi^2 \right)
\end{eqnarray}

\textbf{2. Scalar-field equation:}
\begin{eqnarray}
\nabla_a \nabla^a \Phi - m_s^2 \Phi = 0 . \label{scalar-eq}
\end{eqnarray}

\subsection{Static solutions}
The Einstein field equations (\ref{einstein-eq}) admits cylindrically symmetric solution or ``dRGT black string'' in the absence of scalar field, i.e., $T^{\Phi}_{\mu\nu}=0$. The dRGT black string solution is defined as \cite{Tannukij:2017jtn}

\begin{align}
ds^{2} &= -f(r)dt^2 + f^{-1}dr^2 + r^2\alpha_g^2dz^2 + r^2 d\varphi^2, \label{BSmetric}
\end{align}
where 
\begin{align}
f(r) &= \alpha_m^2 r^2 - \frac{4M}{\alpha_g r} + \gamma r + \epsilon, \label{metric} \\
\alpha_m^2 &= m^2_g(1+\alpha+\beta)\equiv-\frac{\Lambda}{3}, \\
\gamma &= -\frac{\alpha_m^2h_0(1+2\alpha+3\beta)}{1+\alpha+\beta}, \\
\epsilon &= \frac{\alpha_m^2h_0^2(\alpha+3\beta)}{1+\alpha+\beta}.
\end{align}
$M$ is mass per unit length in $z$ direction of black string. One can clearly see that the graviton mass $m_g$ naturally generates the effect of cosmological constant. With these definitions, $\Lambda<0$ case associates with $\alpha_m^2>0$ while $\Lambda>0$ is obtained by letting $\alpha_m^2<0$. This is the unique character of black string in dRGT massive gravity. Since there is no de-Sitter (dS) analogue of black string in standard four dimensional general relativity \cite{Lemos:1994xp}. The presence of linear term $\gamma$ and constant term $\epsilon$ makes black string with positive $\Lambda$ possible. As will be shown later, these two branches of black string solutions have different horizon and asymptotic structures. Therefore, to determine their QNMs, we need to consider each branch of solution separately. Note that, in the limit, $\gamma=0,\epsilon=0,\alpha_g=\alpha_m$, this metric (\ref{BSmetric}) reduces to four dimensional black string solution in general relativity found by Lemos \cite{Lemos:1994xp}.

\begin{figure}[h]
\centering
 \includegraphics[width=0.6\textwidth]{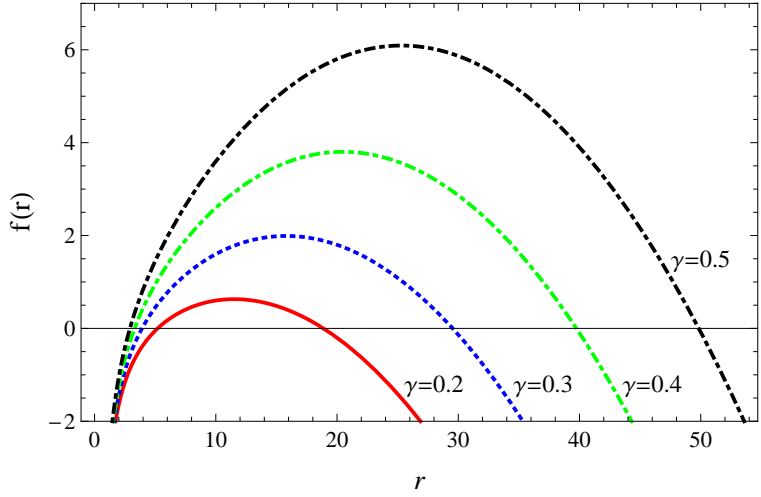}
\caption{ The behaviour of metric function $f(r)$ plotted against radial coordinate $r$ for varying $\gamma$ with $\alpha_m^2=-0.01,M=1,\alpha_g=1,\epsilon=0$. }
\label{fig:dsfig1}
\end{figure}

We shall now investigate the root structure of dRGT black string with negative $\alpha_m^2$. Generally speaking, the metric function (\ref{metric}) has three zeros. With a proper parameter choice, it is possible that all three roots will be real number. More specifically, for $\alpha_m^2<0$ there are two positive real roots and one negative real root. These two positive roots will be treated as black string's event horizon $r_h$ and cosmological horizon $r_c$, where $r_h<r_c$. As an example, the roots structure of metric (\ref{metric}) is displayed in Fig.~\ref{fig:dsfig1}. In this plot, the black string mass $M$, cosmological constant $\alpha_m^2$, $\alpha_g$ and $\epsilon$ are fixed to be $M=1,\alpha_m^2=-0.01,\alpha_g=1,\epsilon=0$ respectively. Each curve represents different values of $\gamma$. These black strings shown in this figure have two positive roots. The inner root is black string's event horizon and the outer root is cosmic horizon. As $\gamma$ increases, the cosmic horizon increases, but the event horizon slightly decreases. In our previous work \cite{Burikham:2017gdm}, we showed that the charged dRGT black hole has different properties ranging from naked singularity to a black hole and extremal black hole. However, this is not the case for neutral black string with positive $\Lambda$. Without the charge term, the metric (\ref{metric}) cannot develop a naked singularity or extremal scenario (for charged black string~\cite{Tannukij:2017jtn}, naked singularity can exist just like in the black hole case). Note that throughout this work, $\epsilon$ will be fixed to zero \cite{Burikham:2017gdm}. However, this setting prevents us from having black string solution with negative and vanishing $\gamma$.

\begin{figure}[h]
\centering
 \includegraphics[width=0.6\textwidth]{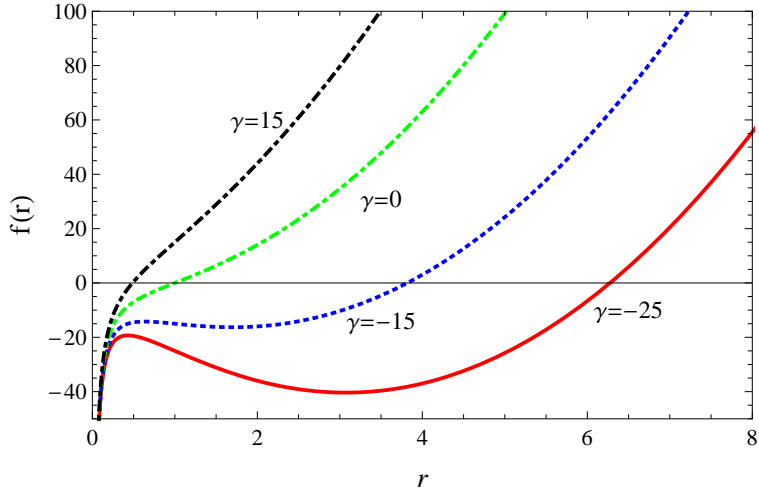}
\caption{ The behaviour of metric function $f(r)$ plotted against radial coordinate $r$ with differing $\gamma$. For demonstration, we fix $\alpha_m^2=4,M=1,\alpha_g=1,\epsilon=0$. }
\label{fig:adsfig1}
\end{figure}

An example of black string solution with $\alpha_m^2>0~(\Lambda<0)$ can be found in Fig.~\ref{fig:adsfig1}. In this plot, we set the black string mass, cosmological constant and other constant to be $M=1,\alpha_m^2=4,\alpha_g=1,\epsilon=0$ and consider the effect of $\gamma$ on the spacetime. As can be seen from the plot, the horizon structure dramatically changes from the positive $\Lambda$ case. There is only one positive real root, i.e., black string's event horizon, for vanishing $\epsilon$ \cite{1stfootnote}. Unlike the positive $\Lambda$ case, the black string exists with all possible value of $\gamma$ (negative, zero and positive). We can see that the black string becomes large when $\gamma$ is more negative. It is clear from the metric (\ref{metric}) that, the black string will always have at least one horizon. This is because the positive $\alpha_m^2$ term will be dominant at large $r$. Without charges, one should not expect naked singularity scenario in this case as well.

\section{Scalar perturbations on dRGT black string}\label{sec:pert}

\begin{figure}[h]
\centering
\includegraphics[width=0.46\textwidth]{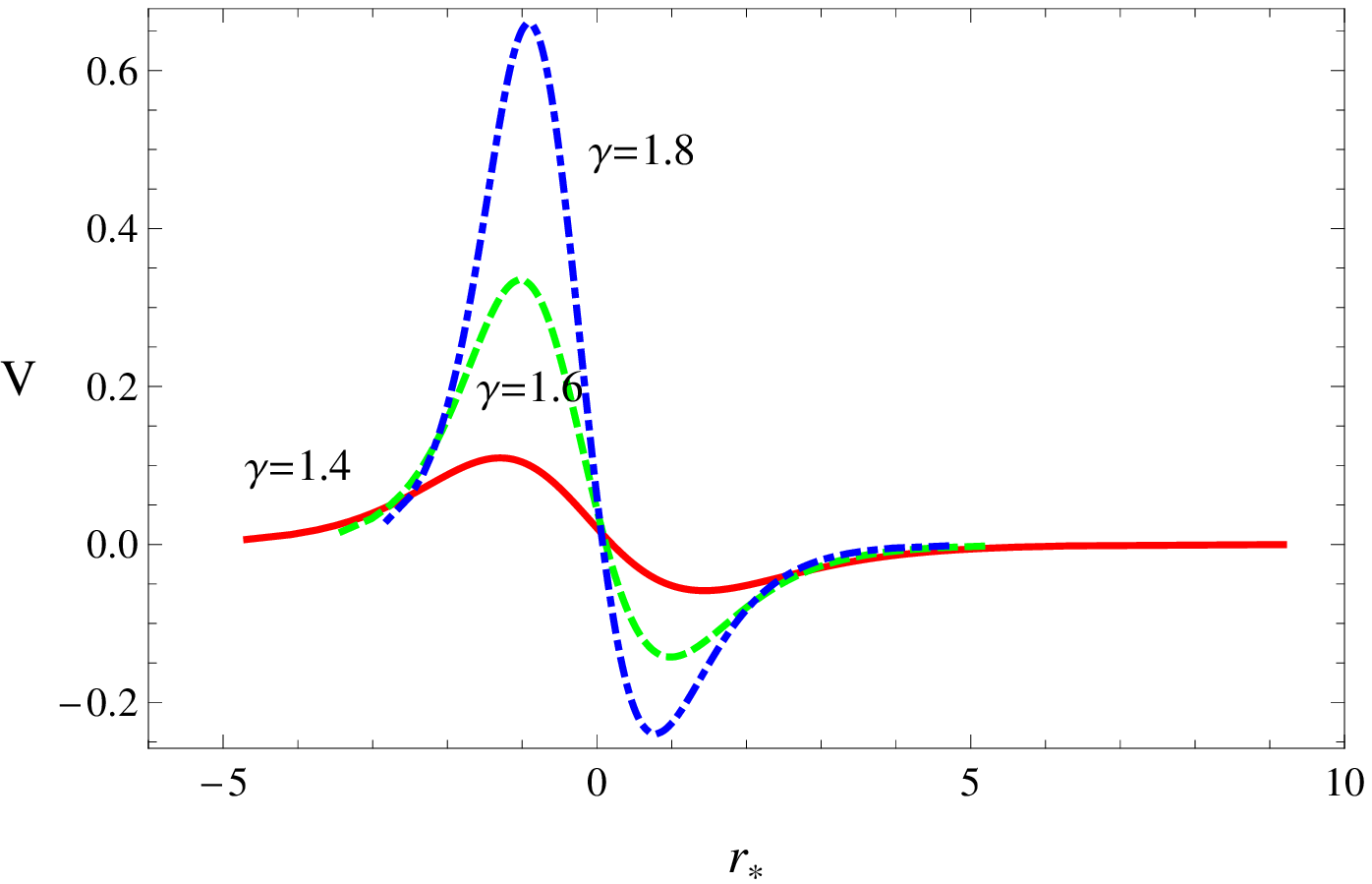}
\includegraphics[width=0.46\textwidth]{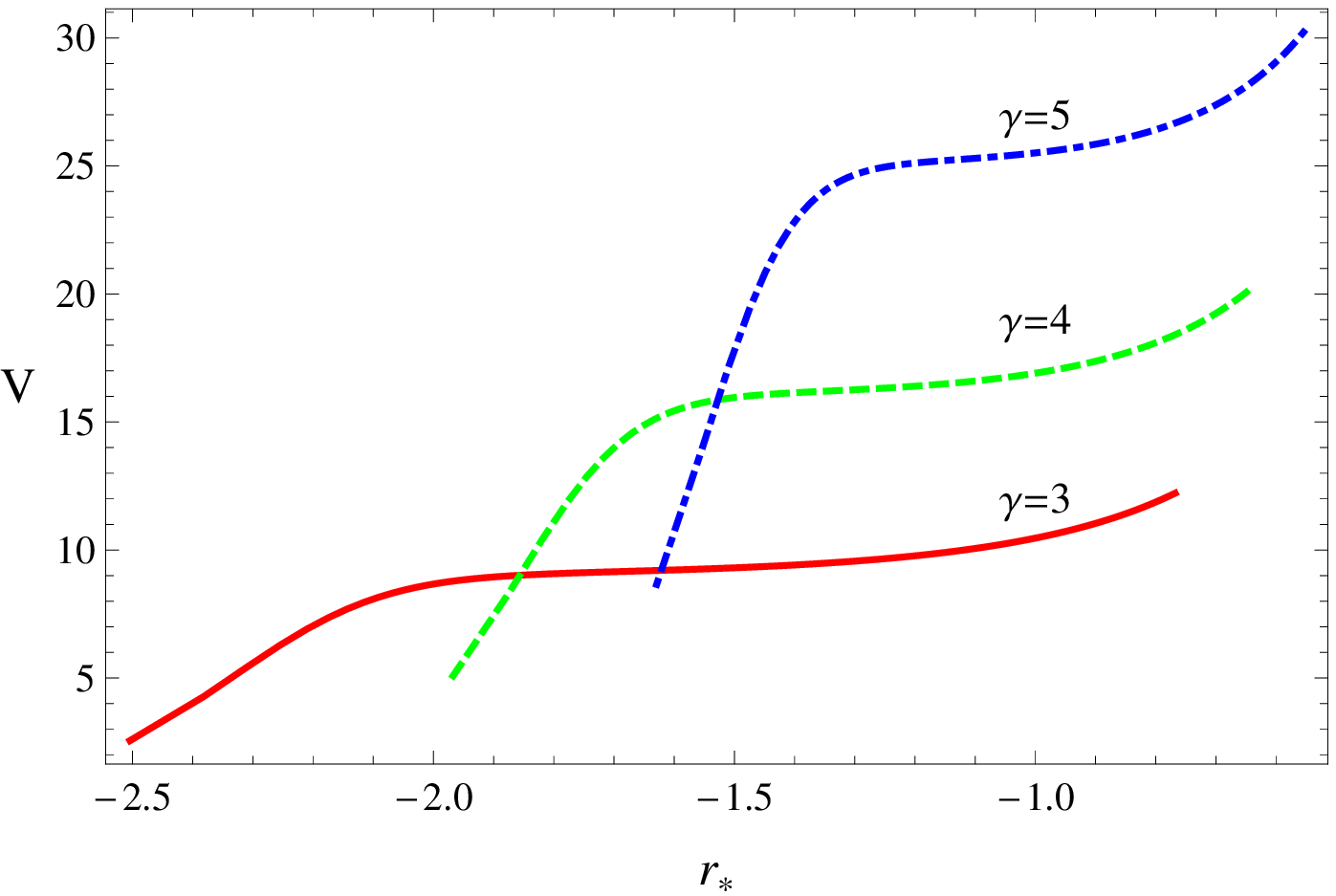}
\caption{The effective potential plotted against tortoise coordinate $r_\ast$ for $M=1,\alpha_g=1,\epsilon=0$ with varying $\gamma$. (Left) $\alpha_m^2=-0.25,\lambda=k=m_s=0$ (Right) $\alpha_m^2=0.003,\lambda=1,k=0,m_s=0.03$.}
\label{fig:veff}
\end{figure}

Now we shall consider a massive scalar field as a test field on dRGT black string background (\ref{BSmetric}). This can be described by the Klein-Gordon equation (\ref{scalar-eq}). We make the following ansatz for the scalar field $\Phi$,

\begin{align}
\Phi &= \frac{\phi(r)}{r}e^{-i\omega t+ikz+i\lambda\varphi},
\end{align}
where $\omega,k,\lambda$ are the frequency, the wave number and angular quantum number of the scalar perturbation. Unlike spherically symmetric set-up where we normally obtain (decoupled) radial and angular part of the Klein-Gordon equation, in this case, however, we solely get a second order ordinary differential equation

\begin{align}
\frac{f}{r} \phi'' + \frac{f'}{r} \phi' + \left(\frac{\omega^2}{r f}-\frac{m_s^2}{r}-\frac{f'}{r^2}-\frac{\lambda^2}{r^3}-\frac{k^2}{\alpha_g^2r^3}\right)\phi &= 0, \label{KG}
\end{align}

where $f'=df/dr$. This equation (\ref{KG}) can be recast into the Schr\"{o}dinger-like equation

\begin{align}
\frac{d^2\phi}{dr_\ast^2} + \left[\omega^2 - V(r)\right]\phi &= 0, \label{KGtortoise}
\end{align}
where the effective potential is given by
\begin{align}
V(r) &= f\left( m_s^2 + \frac{\lambda^2}{r^2} +\frac{k^2}{\alpha_g^2 r^2} + \frac{f'}{r}     \right).
\end{align}
We have introduced the tortoise coordinate $r_\ast$
\begin{align}
\frac{dr_\ast}{dr} &= \frac{1}{f}, \label{tortoise}
\end{align}
with $-\infty<r_\ast<0$, where $r_\ast\rightarrow-\infty$ near the event horizon and $r_\ast\rightarrow0$ at infinity if $\alpha_m^2>0$. For $\alpha_m^2<0$, $r_\ast\to-\infty,\infty$ as $r$ approaches black string's event horizon $r_h$ and cosmological horizon $r_c$, respectively. Note that, in the massless limit $m_s\to0$, equation (\ref{KGtortoise}) resembles those found in \cite{Cardoso:2001vs}. In Fig.~\ref{fig:veff}, we display the effective potentials for positive and negative $\alpha_m^2$. In both cases, the effective potentials increase with $\gamma$. For $\alpha_m^2<0$, the potential vanishes at the event horizon and cosmic horizon. In contrast, the potential increases with $r_\ast$ and diverges as $r_\ast\to 0$ when $\alpha_m^2>0$. 

The main task of this paper is to solve the radial equation (\ref{KGtortoise}) with appropriate boundary conditions. If $\alpha_m^2<0$, effective potential vanishes at both ends~(see Fig.~\ref{fig:veff}), i.e., the event and cosmic horizon. At these points, two independent solutions are $\phi\sim e^{\pm i\omega r_\ast}$. With the presence of horizons, only ingoing modes are allowed at the event horizon $\phi_{r_h}\sim e^{-i\omega r_\ast}$ and outgoing modes at the cosmic horizon $\phi_{r_c}\sim e^{i\omega r_\ast}$. For $\alpha_m^2>0$ case, the condition at the event horizon is still the same. However, since the effective potential diverges as $r$ approaches infinity, we require the wave function to vanish at infinity. To satisfy these quasinormal boundary conditions, the associated quasinormal frequencies $\omega$ will be discrete complex numbers. With the sign convention used in this work, scalar perturbations will be stable if Im$(\omega)<0$ (decaying) and unstable if Im$(\omega)>0$ (growing).

Since each branches of solution has different boundary condition, we shall apply two distinct numerical methods for calculating the QNMs. For positive $\Lambda$ branch, we implement the so-called asymptotic iteration method (AIM). This method was firstly used to solve an eigenvalue problem \cite{AIM:2003}. An improve version of AIM was developed and applied to compute the QNMs of asymptotically flat Schwarzschild and Schwarzshild de-Sitter black holes \cite{Cho:2009cj}. Recently, the QNMs of black holes in dRGT massive gravity \cite{Prasia:2016fcc,Burikham:2017gdm} and Lifshitz gravity \cite{Zangeneh:2017rhc} are calculated using improved AIM technique. When $\Lambda$ is negative, the spectral method~(see Ref.~\cite{Bonazzola:1998qx} and references therein) is used to numerically calculate the QNMs. This method is recently used for computing the QNMs of charged dRGT black hole in asymptotically AdS space \cite{Burikham:2017gdm}.

\section{Near extremal dRGT black string: $\alpha_m^2<0$}\label{sec:nearEx}

One can obtain an analytic formula for the quasinormal frequency, if the effective potential in equation (\ref{KGtortoise}) can be written in some well-known form. This method was suggested by Ferrari and Mashhoon in 1984 \cite{Ferrari:1984zz}.  It turns out that our potential can be put into those exact forms by considering the near extremal limit. By near extreme, we mean the spacetime which the event horizon $r_h$ is very close to the cosmic horizon $r_c$, i.e., $\displaystyle{\frac{r_c-r_h}{r_c}\ll 1}$  \cite{Cardoso:2003sw}. Therefore in this section, we shall compute the QNMs of dRGT black strings with $\alpha_m^2<0$ in the near extremal limit. 

In general, the metric function (\ref{metric}) has three distinct roots at $r_h$, $r_c$ and $r_0$. They can be written in the following form
\begin{align}
f(r) &= \frac{\alpha_m^2}{r}(r-r_0)(r-r_h)(r-r_c).
\end{align}
We define surface gravity associated to the event horizon 
\begin{align}
\kappa_h &\equiv \left. \frac{1}{2}\frac{df}{dr} \right|_{r=r_h}, \nonumber \\
		 &= \frac{\alpha_m^2(r_h-r_0)(r_h-r_c)}{2r_h}.
\end{align}
Three zeroes of $f$ can be expressed in term of $\alpha_m^2,\gamma,\epsilon$ and $M$ as follow
\begin{align}
r_0+ r_h + r_c &= -\frac{\gamma}{\alpha_m^2}, \\
r_h^2 + r_hr_c + r_c^2 + \left(r_h + r_c\right)\frac{\gamma}{\alpha_m^2} &= -\frac{\epsilon}{\alpha_m^2}, \\
r_hr_c\left(r_h+r_c\right)+\frac{\gamma}{\alpha_m^2}r_hr_c &= -\frac{4M}{\alpha_g\alpha_m^2}.
\end{align}
In the near extremal limit, we shall consider the following approximation
\begin{align}
r_0 &\sim -\left[2r_h + \frac{\gamma}{\alpha_m^2}\right], \\
\alpha_m^2 &\sim -\left[\frac{2\gamma}{3r_h}+\frac{\epsilon}{3r_h^2}\right], \\
M &\sim \frac{\alpha_gr_h}{6}\left[\epsilon+\frac{\gamma r_h}{2}\right], \\
\kappa_h &\sim -\frac{(r_h-r_c)}{2r_h^2}\left[\gamma r_h + \epsilon\right].
\end{align}
Since $r$ varies between $r_h$ and $r_c$, thus $r-r_0\sim r_h-r_0$. Therefore we can write down the metric function in the near extremal limit as
\begin{align}
f &\sim \frac{\alpha_m^2}{r_h}(r_h-r_0)(r-r_h)(r-r_c), \nonumber \\
  &\sim -\frac{(r-r_h)(r-r_c)}{r_h^2}\left[\gamma r_h + \epsilon\right]. \label{metricapprox}
\end{align}
With the definition of the tortoise coordinate (\ref{tortoise}), this allows us to express the radial coordinate $r$ as a function of $r_\ast$. Now it reads,
\begin{align}
r &= \frac{r_c e^{2\kappa_h r_\ast}+ r_h}{1+e^{2\kappa_h r_\ast}}.
\end{align}
Thus the metric function (\ref{metricapprox}) can be expressed in the tortoise coordinate as
\begin{align}
f &\sim \frac{\left(r_h\kappa_h\right)^2}{\left(\gamma r_h+\epsilon\right)\cosh^2(\kappa_h r_\ast)}.
\end{align}
By the virtue of these approximations, the Klein-Gordon equation (\ref{KGtortoise}) in the near extremal limit becomes
\begin{align}
\frac{d^2\phi}{dr_\ast^2} + \left[\omega^2 - \frac{V_0}{\cosh^2(\kappa_h r_\ast)} \right]\phi &= 0, \label{KGtortoiseapprox}
\end{align}
where
\begin{align}
V_0 &= \frac{\kappa_h^2}{\left(\gamma r_h + \epsilon\right)}\left[m_s^2 r_h^2 + \lambda^2 + \frac{k^2}{\alpha_g^2}\right].
\end{align}
This potential is the well-known P\"oschl-Teller potential \cite{Poschl1933}. The bound states of this type of potential were studied extensively in \cite{Poschl1933}. The P\"oschl-Teller technique was used to investigate the QNMs of near extremal Schwarzschild-dS \cite{Cardoso:2003sw,Zhidenko:2003wq}, $d$ dimensional Reissner-Nordstr\"om-dS black hole \cite{Molina:2003ff} and black hole in massive gravity \cite{Fernando:2014gda}. 

When one applies the boundary conditions of QNMs to (\ref{KGtortoiseapprox}), the associated quasinormal frequency $\omega_n$ is given by \cite{Ferrari:1984zz,Cardoso:2003sw}
\begin{align}
\omega_n &= \kappa_h\left[\sqrt{\frac{V_0}{\kappa_h^2}-\frac{1}{4}} -\left(n+\frac{1}{2}\right)i\right],~~~~~~~~~~~~~n=0,1,2... \label{wformula}
\end{align}
It is clear that the quasinormal frequency is quantized by an integer $n$. The most fundamental mode is characterized by $n=0$, i.e., the most slowly decaying modes. As an example, the first three lowest modes of the quasinormal frequencies are shown in Table~\ref{tab:dstab1}. In this table, we fix the mass and other parameters to be $M=1,\alpha_g=1,\epsilon=0,\lambda=0,k=0$. At each mode $n$, the $\omega_n$ is compared between the massless and massive scalar case. From the formula, it should not be surprised that the quasinormal frequencies are purely imaginary for the massless scalar case. It is clear from Table~\ref{tab:dstab1} that increasing mode number $n$ affects only imaginary part of the frequencies.  Comparing to the black hole case studied in Ref.~\cite{Konoplya:2004wg,Konoplya:2006br}, the behaviour of black string QNMs is quite different.  As the scalar mass increases, the imaginary part of the $\omega$ is not decreasing to the quasi-resonant modes as found in the black holes.

We remark that, the formula (\ref{wformula}) is reliable only when the black string's event horizon is very close to the cosmological horizon, $\displaystyle{\frac{r_c-r_h}{r_c}\ll 1}$. In the appropriate limit \cite{2ndfootnote}, equation (\ref{wformula}) resembles those found in \cite{Cardoso:2003sw,Zhidenko:2003wq}.

\setlength{\tabcolsep}{4pt}
\begin{table}[H]
\begin{center}
  \begin{tabular}{|c|c|c|c|c|c|c|}
\hline
\multirow{2}{*}{$\{\alpha_m^2,\gamma\}$} &  \multicolumn{2}{c|}{$\omega_0$} &  \multicolumn{2}{c|}{$\omega_1$}  & \multicolumn{2}{c|}{$\omega_2$}\\ 
\cline{2-7}
&  $m_s=0$  &  $m_s=0.5$ & $m_s=0$   &  $m_s=0.5$ & $m_s=0$   &  $m_s=0.5$ \\\hline
$\{-0.09,0.6025\}$
&$0$ & $0.003165-0.001253i$
&$-0.002507i$ & $0.003165-0.003760i$
&$-0.005014i$ & $0.003165-0.006267i$ \\
\hline 
$\{-0.25,1.1906\}$
&$0$ & $0.004947-0.003852i$
&$-0.007704i$ & $0.004947-0.011556i$
&$-0.015408i$ & $0.004947-0.019260i$
\\ \hline
$\{-0.49,1.8646\}$
&$0$ & $0.000518-0.000863i$
&$-0.001727i$ & $0.000518-0.002590i$
&$-0.003454i$ & $0.000518-0.004317i$
\\
\cline{4-5}
\hline 
 \end{tabular}
    \caption{The QNMs of dRGT black strings with $\alpha_m^2<0$ in the near extremal limit from formula (\ref{wformula}). For demonstration purpose, we set $M=1,\alpha_g=1,\epsilon=0,\lambda=0,k=0$.}
\label{tab:dstab1}
\end{center}
\end{table}

\section{QNMs of dRGT black strings with positive $\Lambda$: Improved AIM}\label{sec:dS}

In this section, we compute the QNMs of black strings in dRGT massive gravity with $\Lambda>0$ by using improved asymptotic iteration method \cite{Cho:2009cj}. To do this, we shall introduce the radial coordinate transformation $r=1/x$. The radial wave equation (\ref{KGtortoise}) becomes
\begin{align}
\phi'' + \frac{A'}{A}\phi' + \left[\frac{\omega^2}{A^2} - \frac{1}{A}\left(\lambda^2+\frac{k^2}{\alpha_g^2}+\frac{m_s^2}{x^2} - xf'\right)\right]\phi &= 0, \label{weqinX}
\end{align}
where
\begin{align}
A(x) &= \alpha_m^2 - \frac{4Mx^3}{\alpha_g} + \gamma x + \epsilon x^2.
\end{align}
Throughout this section, we shall denote derivative with respect to radial coordinate $x$ by $^\prime$. Next we define the following \cite{Prasia:2016fcc,Cho:2009cj,Moss:2001ga}
\begin{align}
e^{i\omega r_\ast} &= \left(x-x_1\right)^{\frac{i\omega}{2\kappa_1}}\left(x-x_2\right)^{\frac{i\omega}{2\kappa_2}}\left(x-x_3\right)^{\frac{i\omega}{2\kappa_3}},
\end{align}
where $x_i$ is real root of $f(r)$. The surface gravity associated with each horizon is given by $\displaystyle{\kappa_i = \frac{1}{2}\frac{df}{dr}\Big|_{r=r_i}}$. In this section, the black string event horizon and cosmological horizon are defined by $x_1$ and $x_2$ whereas the negative real root is $x_3$. To deal with the singularity at the cosmic horizon, we make the following substitution
\begin{align}
\phi &= e^{i\omega r_\ast}u(x).
\end{align}
The wave equation (\ref{weqinX}) now reads
\begin{align}
Au'' + \left(A'-2i\omega\right)u' - \left[\lambda^2+\frac{k^2}{\alpha_g^2}+\frac{4Mx}{\alpha_g}+\frac{\gamma}{x}+\frac{\left(2\alpha_m^2+m_s^2\right)}{x^2}\right]u &=0.
\end{align}
At the black string's horizon, the divergent behaviour is scaled out by using 
\begin{align}
u(x) &= \left(x-x_1\right)^{-\frac{i\omega}{\kappa_1}}\chi(x).
\end{align}
The final equation becomes
\begin{align}
\chi''(x) &= \lambda_0(x,\omega)\chi'(x) + s_0(x,\omega)\chi(x), \label{finaleq}\\
\lambda_0 &= \frac{i\alpha_g \omega }{M\left(x-x_1\right)\left(x_1-x_2\right)\left(x_1-x_3\right)} - \frac{\left(A'-2i\omega\right)}{A}, \label{lambda0} \\
s_0 &= \frac{1}{A}\left[\lambda^2+\frac{k^2}{\alpha_g^2}+\frac{4Mx}{\alpha_g}+\frac{2\alpha_m^2+m_s^2+\gamma x}{x^2} + \frac{\alpha_g\omega^2}{M\left(x-x_1\right)\left(x_1-x_2\right)\left(x_1-x_3\right)}\right] \nonumber \\
 &~~~+ \frac{i\alpha_g\omega f'}{2Mf\left(x-x_1\right)\left(x_1-x_2\right)\left(x_1-x_3\right)} + \frac{\alpha_g\omega\left(2iM\left(x-2x_1\right)\left(x_1-x_2\right)\left(x_1-x_3\right)+\alpha_g\omega x\right)}{4M^2x\left(x-x_1\right)^2\left(x_1-x_2\right)^2\left(x_1-x_3\right)^2}. \label{s0}
\end{align}

Equation (\ref{finaleq}) together with the coefficients (\ref{lambda0}), (\ref{s0}) form a core behind an improved AIM algorithm. We choose to omit the details of the numerical method used in this paper. For interested readers, we refer to Refs~\cite{Cho:2009cj,Prasia:2016fcc} for more details on AIM.

\subsection{Results}  \label{numre}

\begin{figure}[h]
\centering
 \includegraphics[width=0.7\textwidth]{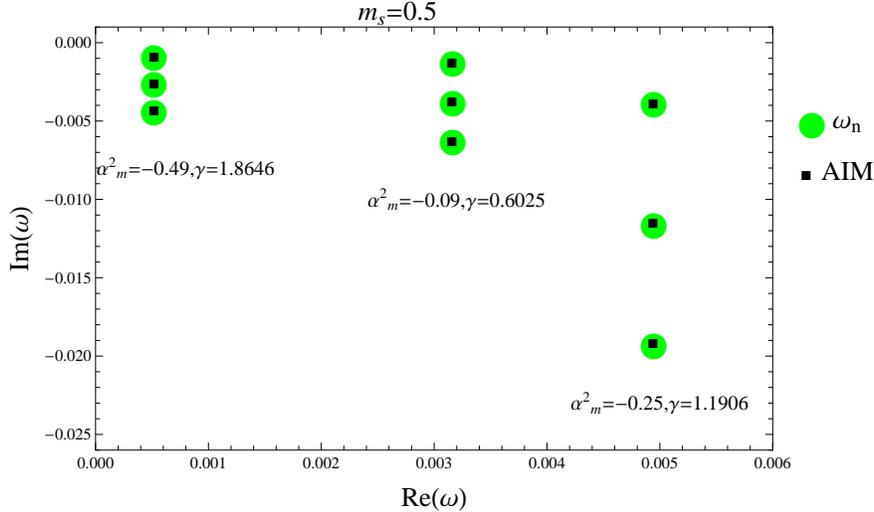}
\caption{The comparison between the QNMs in the near extremal limit obtained via formula (\ref{wformula}) and numerical method of AIM.}
\label{fig:dsfig2}
\end{figure}

The QNMs of the dRGT black strings with $\alpha_m^2<0~(\Lambda>0)$ are calculated using Mathematica notebook modified
from the one provided in Ref \cite{WadeNB}. As a preliminary check, the results from Table~\ref{tab:dstab1} (massive scalar case) are reproduced by using an improved AIM. The first three lowest modes are then compared between the analytic expression (\ref{wformula}) and AIM. The results are in a good agreement as displayed in Fig.~\ref{fig:dsfig2}. Note that we shall focus only on the modes with lowest imaginary parts for most situations.

Now we turn our attention to a non-extremal case. In Table~\ref{tab:dstab2}, we compute the QNMs for massive scalar perturbations for various values of $\gamma$. At each fixed $\gamma$, the value of angular quantum number $\lambda$ is varied from $0-2$. We observe that as $\gamma$ increases, the real part and imaginary part of the quasinormal frequencies also increase (in magnitude for Im$(\omega)$). An increasing in Re$(\omega)$ corresponds to a more oscillation of the scalar wave which means it has more energy. From Im($\omega$), it turns out that the mode with high energy decays faster than the low energy mode. This is because as $\gamma$ gets larger gravity becomes more attractive. Stronger gravity implies that the scalar wave tends to decay faster. We also see that, Re$(\omega)$ increases as angular quantum number increases while Im$(\omega)$ slightly decreases (in magnitude). In the fourth column, the quasinormal frequencies from the third-order WKB approximation (see appendix of Ref.~\cite{Burikham:2017gdm} for details, see also a more accurate sixth-order WKB method in Ref. \cite{Konoplya:2003ii} and the semi-analytic method in Ref. \cite{Matyjasek:2017psv}) are shown. The results show a good agreement between the two methods (AIM and WKB).

\setlength{\tabcolsep}{15pt}
\begin{table}[h]

\centering

\begin{tabular}{|c|c|c|c|c|}
\hline
\multicolumn{4}{|c|}{The QNMs calculated from AIM (50 iterations)}&Third order WKB \\
\hline
$\gamma$ & $\omega(\lambda=0)$ & $\omega(\lambda=1)$ & $\omega(\lambda=2)$  & $\omega(\lambda=2$) \\  \hline

$1.2$ & 0.069001$-$0.053417$i$  & 0.088038$-$0.053369$i$ & 0.129259$-$0.053230$i$ & 0.129240$-$0.053434$i$   \\
 
\hline
 
$1.4$ & 0.382962$-$0.284234$i$  & 0.466561$-$0.279395$i$  & 0.653469$-$0.273342$i$ & 0.655922$-$0.275305$i$   \\
 
\hline

$1.6$ & 0.619823$-$0.424720$i$  & 0.726152$-$0.415527$i$ & 0.974767$-$0.403780$i$  & 0.979613$-$0.403648$i$ \\
 
\hline

$1.8$ & 0.850808$-$0.535069$i$  & 0.969931$-$0.524651$i$ & 1.259101$-$0.510741$i$ & 1.262457$-$0.506438$i$  \\
 
\hline

$2.0$ & 1.078052$-$0.626748$i$  & 1.206173$-$0.616840$i$ & 1.525824$-$0.603461$i$ & 1.524587$-$0.594328$i$  \\
 
\hline 
\end{tabular}
\caption{The QNMs for massive scalar perturbations of a neutral black string for $M=1,\alpha_g=1,\epsilon=0,\alpha_m^2=-0.25,m_s=0.5,k=0$. }
\label{tab:dstab2}
\end{table}

The effect of $\alpha_m^2$ on the QNMs is investigated in Table~\ref{tab:dstab3}. Let's remark that as $\alpha_m^2$ becomes more negative the difference between black string's event horizon $x_1$ and cosmic horizon $x_2$ is smaller. When $\alpha_m^2$ is larger (in magnitude), the real part of $\omega$ is smaller for each fixed $\lambda$. However, the behaviour of the imaginary part of $\omega$ is not trivial as $\alpha_m^2$ increases. In our chosen parameters, Im$(\omega)$ increases as $\alpha_m^2$ changes from $-0.16$ to $-0.25$ and then decreases as $\alpha_m^2$ gets larger. For fixed $\lambda=2$, however, Im$(\omega)$ gets lower with increasing $\alpha_m^2$.  We also find that at each value of $\alpha_m^2$, Re$(\omega)$ increases with $\lambda$ whereas Im$(\omega)$ decreases marginally as happens in Table~\ref{tab:dstab2}~(except when $\alpha_{m}^{2}=-0.16$). We find a close agreement between the results from AIM and the WKB approximation.

\setlength{\tabcolsep}{15pt}
\begin{table}

\centering

\begin{tabular}{|c|c|c|c|c|}
\hline
\multicolumn{4}{|c|}{The QNMs calculated from AIM (50 iterations)}&Third order WKB \\
\hline
$\alpha_m^2$ & $\omega(\lambda=0)$ & $\omega(\lambda=1)$ & $\omega(\lambda=2)$  & $\omega(\lambda=2$) \\  \hline

$-0.16$ & 1.360627$-$0.533564$i$  & 1.470631$-$0.540154$i$ & 1.766217$-$0.558983$i$ & 1.745391$-$0.544495$i$   \\
 
\hline
 
$-0.25$ & 0.964854$-$0.582753$i$  & 1.088791$-$0.572458$i$  & 1.394118$-$0.558558$i$ & 1.395506$-$0.551890$i$    \\
 
\hline

$-0.36$ & 0.543604$-$0.521306$i$  & 0.661337$-$0.485114$i$ & 0.935650$-$0.449515$i$ & 0.946266$-$0.449496$i$  \\
 
\hline

$-0.49$ & 0.076211$-$0.136994$i$  & 0.143397$-$0.133901$i$ & 0.251467$-$0.131540$i$ & 0.250929$-$0.132739$i$  \\
 
\hline

\end{tabular}
\caption{The QNMs for massive scalar perturbations of a neutral black string for $M=1,\alpha_g=1,\epsilon=0,\gamma=1.9,m_s=0.5,k=0$}
\label{tab:dstab3}
\end{table}
In Table~\ref{tab:dstab4}, we consider how scalar mass affects the quasinormal frequencies in the presence of the massive gravity effects. The values of scalar mass $m_s$ are varied from $0-0.5$. In this table, the angular quantum number $\lambda$ is set to be unity and we investigate the effect of wave number $k$ on the QNMs instead. One remarkable feature is that there exists the QNMs with zero real part and negative imaginary part, i.e. these modes are purely decaying. For massless scalar perturbations, the purely imaginary frequencies $\omega$ are found for all values of $k$ considered here. At $k=2$, a QNM with nonzero real part appears. However, nonzero real parts are found more abundantly when the scalar mass $m_s=0.25$ and $0.5$. The energy naturally goes up along with increasing $k$ while the modes with larger wave number tend to have a shorter lifetime. When comparing the results with WKB approximation, we find a good match between the two methods. However, only the QNMs with nonzero real parts can be obtained from the WKB method.  This is natural since WKB requires the positive-energy solution to exist for certain region around the minimum of the effective potential.  For example in the massless case, WKB can only find the non-diffusive QNMs for $k=2$.  Interestingly, AIM also yields this value as the first excited mode $(n=1)$, while the most fundamental mode (obtained from AIM) $(n=0)$ is diffusive as shown in the table.

\setlength{\tabcolsep}{15pt}
\begin{table}

\centering

\begin{tabular}{|c|c|c|c|c|}
\hline
\multicolumn{4}{|c|}{The QNMs calculated from AIM (50 iterations)}&Third order WKB \\
\hline
$m_s$ & $\omega(k=0)$ & $\omega(k=1)$ & $\omega(k=2)$ & $\omega(k=2$) \\  

\hline
$0.00$ & $-0.080947i$  & $-0.154397i$ & $-0.339379i$ & $1.180228-0.496008i$   \\
 
$$ & $$ & $$ & $1.187530-0.502348i$ & $$ \\

\hline
 
$0.25$ & $-$0.299634$i$  & $-$0.376619$i$  & 1.219792$-$0.484424$i$  & 1.215106$-$0.477351$i$  \\
 
\hline

$0.50$ & 1.014665$-$0.409856$i$  & 1.101989$-$0.413391$i$ & 1.332695$-$0.421810$i$  & 1.327514$-$0.418380$i$ \\
 
\hline 
\end{tabular}
\caption{The QNMs for scalar perturbations of a neutral black string for $M=1,\alpha_g=1,\epsilon=0,\alpha_m^2=-0.16,\gamma=1.5,\lambda=1$}
\label{tab:dstab4}
\end{table}

\begin{figure}[h]
\centering
 \includegraphics[width=0.7\textwidth]{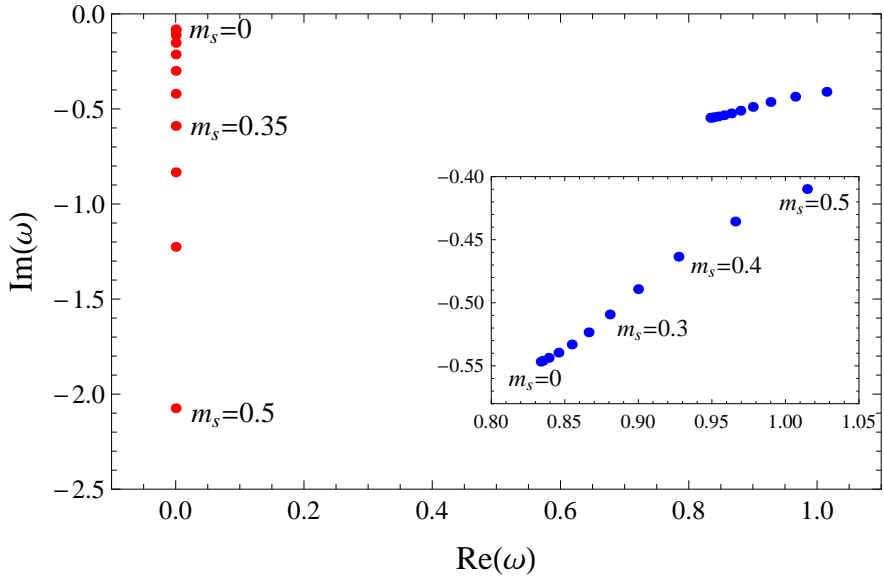}
\caption{The plot of real and imaginary part of quasinormal frequencies for $M=1,\alpha_g=1,\epsilon=0,\alpha_m^2=-0.16,\gamma=1.5,\lambda=1,k=0$ for varying $m_s$. A subplot is displayed to clarify the effect of $m_s$ on the non-zero-real-part modes.
}
\label{fig:dsneatmu1}
\end{figure}

In order to explore the effect of $m_s$ on the peculiar behaviour of the QNMs found in Table~\ref{tab:dstab4}, we plot Re$(\omega)$ and Im$(\omega)$ with smaller increment on the scalar mass to elucidate transit of the lowest mode from the purely imaginary to the values with real parts. This is displayed in Fig.~\ref{fig:dsneatmu1} and the values are listed in Table~\ref{tab:dstab5}. The frequencies plotted in this figure are modes with the lowest imaginary part (in magnitude) for both the purely imaginary mode and the non-zero-real-part mode. The transition of the lowest QNM from purely imaginary mode to non-zero-real-part mode occurs between $m_s=0.3-0.35$. A closer investigation reveals that the transitions appear at $m_s=0.33,0.30,0.21$ for $k=0,1,2$ respectively.

The transition is part of a bigger picture of the QNM pattern which consists of two distinct branches found by our numerical method. The first one is purely imaginary mode where Im$(\omega)$ increases with the scalar field mass. The second branch exists with non-zero real part. In this case, $m_s$ affects the quasinormal frequency such that as the scalar mass increases Re$(\omega)$ also increases but Im$(\omega)$ decreases.  For low $m_{s}$, the non-zero-real-part QNMs have larger $|{\rm Im}(\omega)|$ than the purely imaginary mode, and they exist as higher modes.  At transition $m_{s}$, the imaginary parts of these modes become less than the purely imaginary mode and yet both branches always coexist within the QNM chart.  

It is known that the perturbation by a massive field often leads to purely real frequency or quasi-resonance modes \cite{Konoplya:2004wg,Konoplya:2006br}.  In dRGT model with positive $\Lambda$~(negative $\alpha_{m}^{2}$), the effective potential approaches zero in the asymptotic regions, $V\to 0$ as $r_{*}\to \pm \infty$ as we can see from Figure~\ref{fig:veff}.  This is generic since $V\sim f(r)\sim 0$ at both $r_{h}$ and $r_{c}$.  As a result, the effect of the scalar mass vanishes in the asymptotic regions and there is no quasi-resonance modes.  Our numerical results confirm non-existence of the quasi-resonance modes for positive $\Lambda$ case.

\setlength{\tabcolsep}{15pt}
\begin{table}

\centering

\begin{tabular}{|c|c|c|c|c|}
\hline
\multicolumn{3}{|c|}{The QNMs calculated from AIM (50 iterations)}\\
\hline
$m_s$ & $\omega(k=0)$ & $\omega(k=1)$   \\  

\hline
$0.00$ & $-0.080947i,0.833614-0.546685i$  & $-0.154397i,0.931207-0.531069i$    \\
 

\hline

 $0.05$  & $-0.088556i,0.834988-0.545944i$ & $-0.162217i,0.932553-0.530300i$   \\

\hline

 $0.10$  & $-0.111868i,0.839060-0.543644i$ & $-0.186141i,0.936567-0.527935i$    \\

\hline

 $0.15$  & $-0.152447i,0.845694-0.539523i$ & $-0.227656i,0.943206-0.523776i$    \\

\hline

 $0.20$  & $-0.213303i,0.854735-0.533086i$ & $-0.289567i,0.952440-0.517461i$    \\

\hline
 
$0.25$ & $-0.299634i,0.866175-0.523436i$  & $-0.376619i,0.964399-0.508413i$       \\

\hline

 $0.30$  & $-0.420146i,0.880569-0.509254i$ & $0.979598-0.495876i,-0.496644i$    \\

\hline

 $0.35$  & $0.899789-0.489182i,-0.589268i$ & $0.999254-0.479209i,-0.663064i$  \\

\hline

 $0.40$  & $0.927301-0.463475i,-0.833302i$ & $1.025306-0.458613i,-0.902628i$    \\

\hline

 $0.45$  & $0.965850-0.435581i,-1.225373i$ & $1.059480-0.435813i,-1.288794i$   \\
 
\hline

$0.50$ & $1.014665-0.409856i,-2.074154i$  & $1.101989-0.413391i,-2.118735i$    \\
 
\hline

%
 
\end{tabular}
\caption{The quasi-normal frequencies of the two lowest modes for scalar perturbations of a neutral black string for $M=1,\alpha_g=1,\epsilon=0,\alpha_m^2=-0.16,\gamma=1.5,\lambda=1$ with varying $m_s$.}
\label{tab:dstab5}
\end{table}

\begin{figure}[h]
    \centering
    \begin{minipage}{0.5\textwidth}
        \centering
        \includegraphics[width=0.9\textwidth]{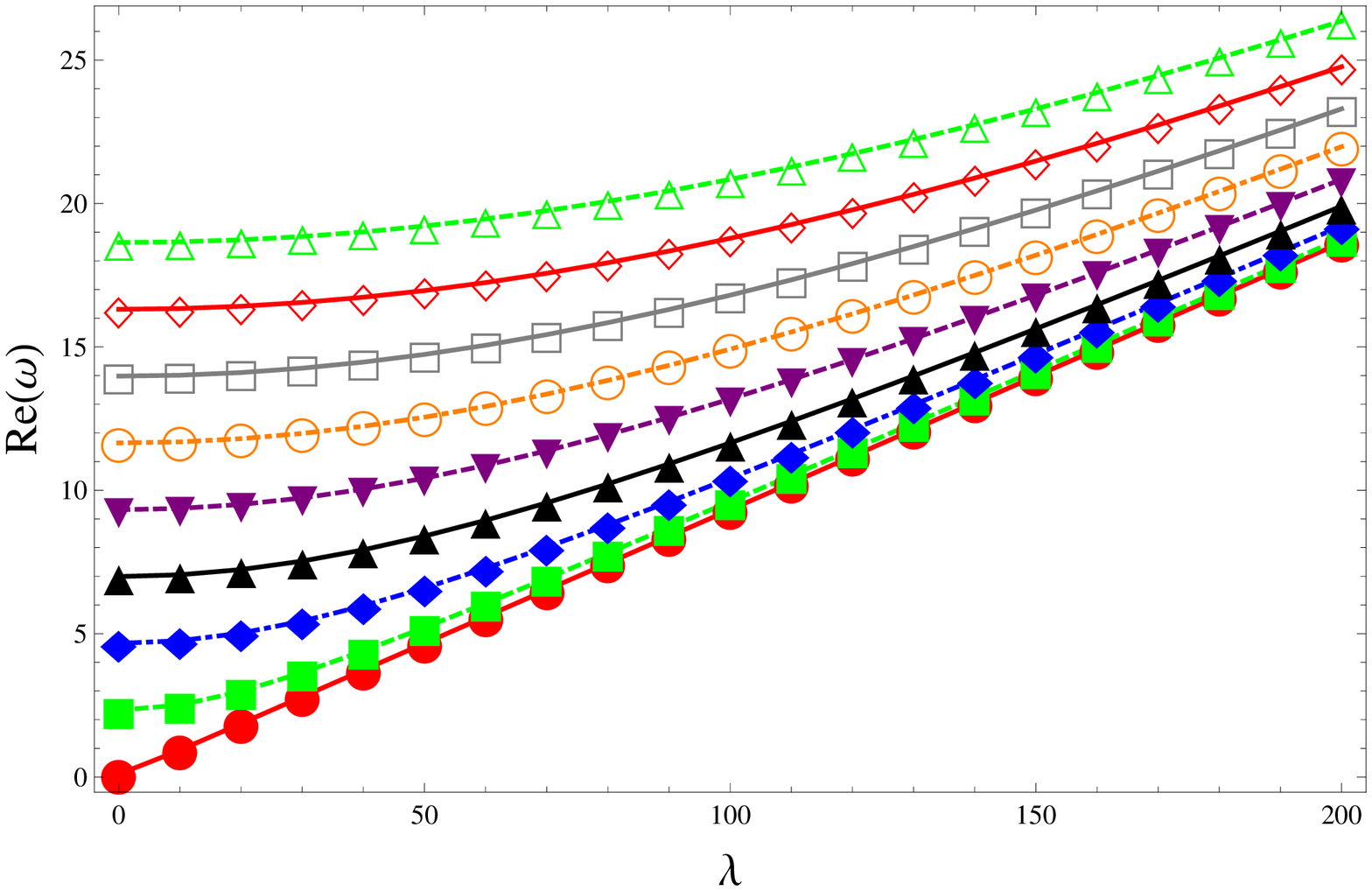} 
    \end{minipage}\hfill
    \begin{minipage}{0.5\textwidth}
        \centering
        \includegraphics[width=1.2\textwidth]{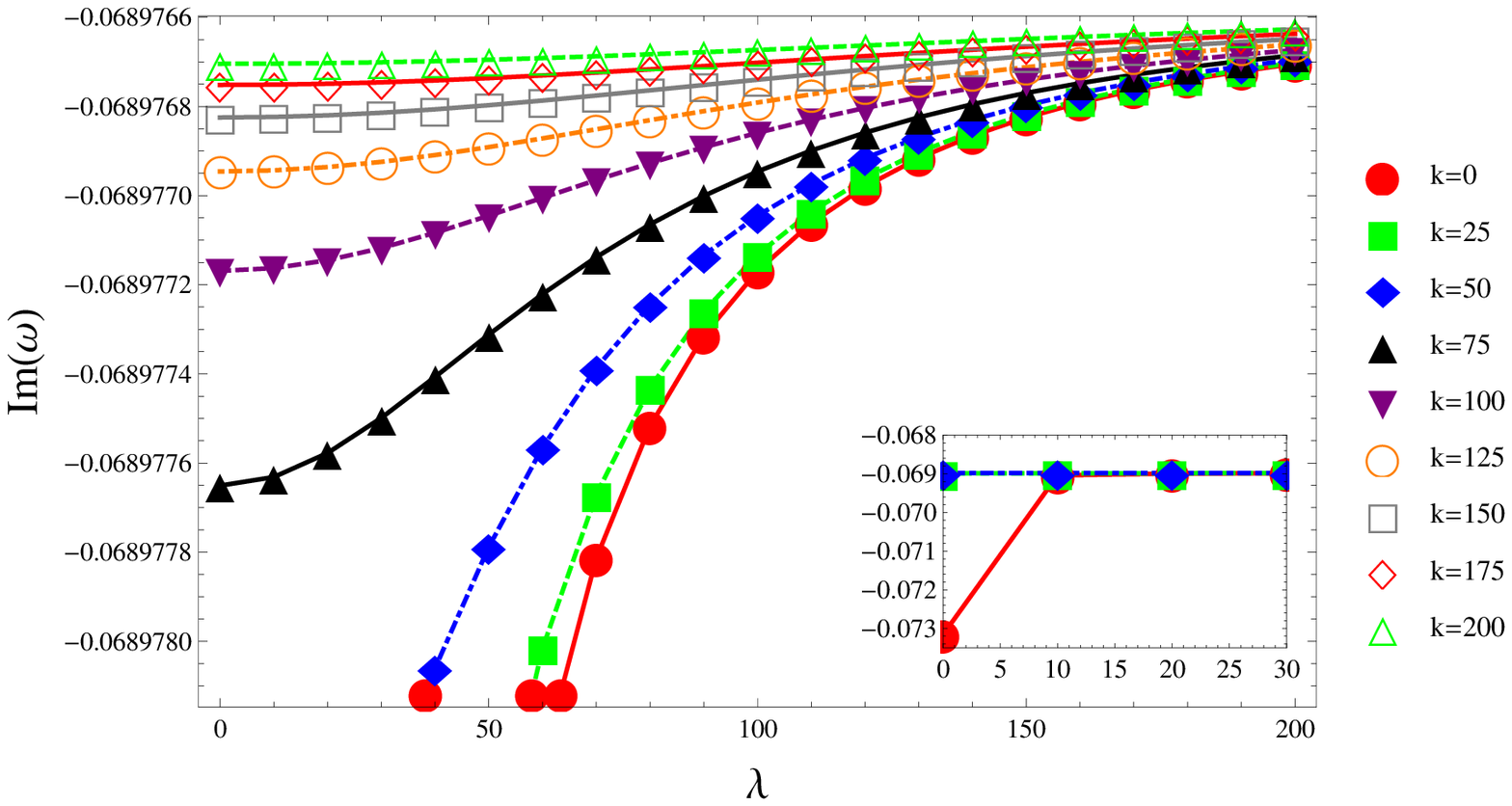} 
    \end{minipage}
    \caption{The behaviour of QNMs at large $\lambda,k$ for $M=1,\alpha_g=1,\epsilon=0,\alpha_m^2=-0.04,\gamma=0.4,m_s=0.2$. (Left) Plot of Re$(\omega)$ as a function of $\lambda$, (Right) Plot of Im$(\omega)$ as a function of $\lambda$, a small window shows the behaviour of $k=0$ curve for small $\lambda$. }
\label{fig:dsfig3}
\end{figure}

One remarkable feature of the QNMs of neutral dRGT black string is the behaviour of the quasinormal frequency at high momentum, $\lambda,k \gg 1$. This is illustrated in Fig.~\ref{fig:dsfig3}. We set $M=1,\alpha_g=1,\epsilon=0,\alpha_m^2=-0.04,\gamma=0.4,m_s=0.2$ and vary $\lambda,k$ from 0 to 200. It is apparent that the scalar wave becomes more energetic as $\lambda$ and/or $k$ get larger. Moreover, the energy of the wave changes rapidly at small wave number $k$. From the plot of Im$(\omega)$, the higher $\lambda,k$ the wave decays slower. Similar to the real part, the effect of $\lambda$ on the QNMs is significant only at the low $k$. We expect the similar pattern if $\lambda$ and $k$ are interchanged in the above figure. This is because from the equation of motion (\ref{KGtortoise}), the roles of $\lambda$ and $k$ to the QNMs are identical up to $1/\alpha_g^2$ factor.

\subsection{WKB approximation for high-momentum modes}\label{wkbhk}

From the first-order WKB approximation, it is possible to estimate the quasinormal frequencies in the large-momentum limit. Note that the following calculations were also done in the Schwarzschild case \cite{Ferrari:1984zz} and the Schwarzschild-dS case \cite{Zhidenko:2003wq}. The first-order WKB approximation yields the following relation \cite{Schutz:1985zz,Iyer:1986np}
\begin{align}
\frac{iQ(r_1)}{\sqrt{2Q^{(2*)}(r_1)}}=\left(n+\frac{1}{2}\right), \label{WKB1st}
\end{align}
where $Q^{(2*)}$ is the second derivative of $Q$ with respect to the tortoise coordinate $r^*$, $r_1$ minimizes $Q(r^*)$ (in terms of the tortoise coordinate), and $Q$ is defined as
\begin{align}
Q\equiv \omega^2-f\left(m^2_s + \frac{\ell^2}{r^2}+\frac{f'}{r}\right),
\end{align}
where $\ell^2\equiv\lambda^2+\displaystyle{\frac{k^2}{\alpha_g^2}}$. With $\epsilon=0,$ in the large-momentum limit $Q$ is then given by 
\begin{align}
Q\approx \omega^2-\frac{f\ell^2}{r^2} = \omega^2 - \frac{\ell^2}{r^2}\left(\alpha^2_m r^2 - \frac{4M}{\alpha_g r} + \gamma r\right).
\end{align}
A local minimum of $Q(r^*)$ can be found as a solution to the following
\begin{align}
Q^{(1*)}(r_1) &= \left.\frac{dQ}{dr^*}\right|_{r_1} = \left.f\frac{dQ}{dr}\right|_{r_1}, \notag
\\
&=-\frac{f(r_{1})\ell^2}{r^4_1}\left(\frac{12M}{\alpha_g}-\gamma r^2_1\right),\notag
\end{align}
The local minimum of $Q(r^*)$ implies either
\begin{align}
f(r_{1}) = 0 \text{ or } \frac{12M}{\alpha_g}-\gamma r^2_1=0. \notag
\end{align}
Since we demand that $r_1$ minimizes $Q$, then we should have $Q^{(2*)}(r_1)>0$. Thus $Q^{(2*)}$ is now calculated
\begin{align}
Q^{(2*)} &= \frac{d^2Q}{(dr^*)^2} = f\frac{d}{dr}\left(f\frac{dQ}{dr}\right),  \notag
\\
&=-f\ell^2\left[f'\left(\frac{12M}{\alpha_g r^4}-\frac{\gamma}{r^2}\right)+f\left(-\frac{48M}{\alpha_g r^5}+\frac{2\gamma}{r^3}\right)\right].  \notag
\end{align}
Since the first choice $f(r_{1})=0$ yields vanishing $Q^{(2*)}(r_1)$, therefore we must choose the second choice which implies $r_1=\sqrt{\frac{12M}{\alpha_g \gamma}}$. Note that $r_1$ is not defined when $\gamma<0$. Thus we obtain
\begin{align}
Q(r_1) &= \omega^2-\ell^2\left(\alpha^2_m+\frac{8M}{\alpha_g r^3_1}\right), \notag
\\
Q^{(2*)}(r_1)&=\frac{2f^2_1\ell^2\gamma}{r^3_1}.  \notag
\end{align}
By requiring that $r_1>r_h$, we can solve (\ref{WKB1st}) for $\omega$ as
\begin{align}
\omega&=\ell\sqrt{\alpha^2_m + \frac{8M}{\alpha_g r^3_1}}\sqrt{1-\frac{2i\sqrt{\gamma r_1}\left(n+\frac{1}{2}\right)}{\ell}}, \notag
\\
&= \sqrt{\alpha^2_m + \frac{8M}{\alpha_g r^3_1}}\left(\ell-i\sqrt{\gamma r_1}\left(n+\frac{1}{2}\right)\right)+\mathcal{O}\left(\frac{1}{\ell}\right)~\text{for large }\ell.  \label{highlw}
\end{align}
It is interesting to note that the above approximation (\ref{highlw}) has an interesting interpretation in the context of an unstable circular orbit of a null geodesic around a massive object \cite{Cardoso:2008bp}. Since the local minimum of $Q$, which is $r_1$, has been proven to also be a radius of a circular null geodesic, then, according to Ref. \cite{Cardoso:2008bp}, the real part of the QNMs in (\ref{highlw}) can be interpreted as multiples of an angular frequency of the corresponding circular null geodesic while the imaginary part indicates instability timescale of the orbit.

In the large-momentum limit, the imaginary part of the lowest QNMs~($n=0$) approaches the asymptotic value
\be
\text{Im}~\omega(\ell\to\infty)=-\frac{1}{2}\left(\alpha^2_m\sqrt{\frac{12M\gamma}{\alpha_{g}}} + \frac{2}{3}\gamma^{2}\right)^{1/2}.
\ee

Numerical results we found in Fig.~\ref{fig:dsfig3} match the values given by (\ref{highlw}) for $n=0$ with errors less than $0.1\%$ for large momentum.

\section{QNMs of dRGT black strings with negative $\Lambda$: Spectral method}\label{sec:ads}
 
In this section, we will explore the QNMs of scalar perturbation in the black string spacetime in the dRGT model with negative $\Lambda$.  Geometrically, this is not an asymptotically global AdS space since we have cylinder instead of sphere.  The boundary topology is cylindrical but otherwise the holographic nature of the space remains the same as the AdS space.  We can understand the general aspects of the perturbations by considering (\ref{KGtortoise}) in two limits, near-horizon and far-away regions.  For the near-horizon region, the equation of motion becomes
\be
\frac{d^{2}\phi}{dr_{*}^{2}}=-\omega^{2}\phi,
\ee
so the solution is simply $\phi(r)\sim e^{\pm i\omega r_{*}}$.  For QNMs calculation, we take the infalling wave $\phi\sim e^{-i\omega r_{*}}$, the perturbation is leaking into the black string horizon.  For the far-away region since $f(r)\simeq \alpha_{m}^{2}r^{2}=-\Lambda r^{2}/3$, Eqn.~(\ref{KGtortoise}) takes the form
\be
\frac{d^{2}\phi}{dr_{*}^{2}}= -\frac{\Lambda}{3}r^{2}\left( m^{2}_{s}-\frac{2\Lambda}{3}\right)\phi,
\ee  
and has the power-law solution $\phi \sim r^{\alpha}$ where 
\be
\alpha = \frac{1}{2}\left( -1\pm\sqrt{9-\frac{12 m_{s}^{2}}{\Lambda}}\right). \label{psol}
\ee
Note that $m^{2}_{s}$ is bounded by $m_{s}^{2}\leq 3\Lambda/4$.  In order for the field to vanish at infinity for the plus sign solution of (\ref{psol}), $m_{s}^{2}\geq 2\Lambda/3$ is required otherwise it will become a non-normalisable configuration from the viewpoint of holographic duality.  Such non-vanishing mode can generate back-reaction to the background and cause transition to other geometry.  However, the minus sign solution always exists as normalisable mode without the lower bound on $m^{2}_{s}$.

In order to numerically calculate the QNMs, we let $\phi = e^{-i\omega r_{*}}S$ to linearise the equation of motion (\ref{KGtortoise}) with respect to $\omega$ and change the coordinate by $u\equiv r_{h}/r$ so that the physical region is $u\in [0,1]$.  The resulting equation of motion is
\be
u^{2}\frac{\partial}{\partial u}\left( f(u) u^{2}\frac{\partial}{\partial u}S(u)\right)+2iw~u^{2}\frac{\partial}{\partial u}S(u) -\Big[ m^{2}+\Big( \lambda^{2} +\frac{k^{2}}{\alpha_{g}^{2}}\Big)u^{2}-f'(u) u^{3}\Big]S(u) = 0, \label{eomn}
\ee
where $w\equiv \omega r_{h}, m\equiv m_{s}r_{h}$ are dimensionless parameters.  

We shall now calculate the QNMs by using the spectral method~(see Ref.~\cite{Bonazzola:1998qx} and references therein).  This is the method of expanding the solution of the quasinormal mode equation with series of orthonormal functions.  Each basis function is regular both at the horizon and the infinity so that the solution satisfies the regular boundary conditions, i.e. only normalizable modes are considered. First expand the solution for positive integer $N$ 
\be
S(u) = \sum_{n=0}^{N}b_{n}T_{n}(2u-1),
\ee
where $T_{n}$ is the Chebyshev polynomials of the first kind.  The larger value of $N$ gives the more accurate approximation of $S(u)$.  Substituting the expansion into the equation of motion (\ref{eomn}), we obtain the linear equation of coefficients $b_{n}$.  By dividing the domain of interest $(2u-1) \in [-1,1]$ into a finite number of grid points and solve the system of linear equations of coefficients $b_{n}$, the quasinormal frequencies will be determined.  In the numerical calculation, we use the Gauss-Lobatto grid points 
\be
u_{k}=\frac{1}{2}\left(1+\cos\left(\frac{k\pi}{N}\right)\right),
\ee
where $k=0,1,..,N$ and solve the generalized eigenvalue problem to obtain the quasinormal frequencies $w$ for a given $N$.  An excellent example Mathematica code of the spectral method for the calculation of the QNMs is given by Yaffe in Ref.~\cite{yaffec}.

\subsection{Results}
   
\begin{figure}[h]
\centering
 \includegraphics[width=0.7\textwidth]{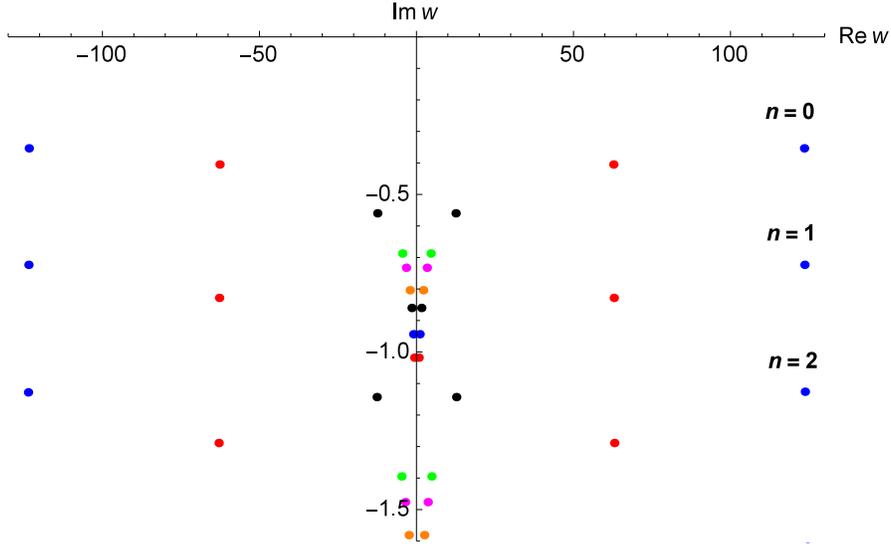}
\caption{The QNMs $w=\omega r_{h}~(r_{h}=10.6266)$ of dRGT black string with $M=1,\alpha_g=1,\epsilon=0,\Lambda=-0.01, \gamma=0, m=0.2$ and $(\lambda, k)=(0,0), (1,0), (2,0), (3,0), (5,0), (7,0), (20,1), (20,100), (20,200)$~(presented by colored dots from the vertical axis out to both sides).  In addition to the lowest modes, the next-to-lowest modes appear for sufficiently high momentum $(\lambda, k)$ in a repeated pattern at exactly the same values of $\text{Re}~w$. }
\label{fignegLamb2}
\end{figure}

For $\gamma =0$ as shown in Table \ref{tadl} for each different angular momentum $\lambda =0, 1, 2$ state at zero $k$ momentum, the lowest lying modes tend to live even longer as the momentum $\lambda$ increases while the energy also naturally increases.  It is apparent from the equation of motion (\ref{KGtortoise}) that the effect of $\lambda$ and $k$ to the QNMs are similar upto the factor of $\alpha_{g}^{2}$ so we expect the similar dependence on $k$.  Another notable feature is the convergence of the lowest QNMs for high momentum $\lambda, k \gg 1$ to a simple power-law asymptotic relation
\be
\text{Im}~w = e^{-\beta}(\text{Re}~w)^{-0.204},  \label{asymp}
\ee
where  e.g. $\beta = 0.08, 0.06, 0.0156$ for $m=0, 0.2, 0.4$ respectively.  This result is consistent with what found in Ref.~\cite{Festuccia:2008zx}~(with $d=3$ modulo topology of the boundary, the exponent $0.204\simeq (d-2)/(d+2)=1/5$) using the WKB method.  Figure \ref{fignegLamb2} shows this asymptotic relation for the lowest modes.  The high momentum limits in both $\lambda$ and $k$ thus give the same asymptotic behaviour governed by (\ref{asymp}).  Naturally for the infinite momentum behaviour, it has a universal power-law for any mass $m_{s}$.  The only effect of $m_{s}$ shows up at low momentum as different values of $\beta$.  These high-momentum modes have $\text{Im}(w)\ll \text{Re}(w)$ and thus can be thought of as quasi-particles with long lifetime in the dual gauge theory.     

\begin{table}[h]
\begin{center}
\begin{tabular}{|c|c|c|}
\hline
$\lambda =0$ & $\lambda =1$ & $\lambda =2$    \\  \hline

$\pm0.704347-1.0179 i$ & $\pm1.01039-0.943509 i$  & $\pm1.54679-0.860185 i$   \\
 
\hline
\end{tabular} 
\caption{The QNMs $w=\omega r_{h}~(r_{h}=10.6266)$ of dRGT black string with $M=1,\alpha_g=1,\epsilon=0,\Lambda=-0.01, \gamma=0, k=0, m=0.2,\lambda=0, 1, 2$. }
\label{tadl}
\end{center}
\end{table}

For positive $\gamma$, the large-momentum behaviour is different in a remarkable way.  Around $\ell =10$~(defined in Section \ref{wkbhk}), some QNMs start to become normal modes with $\text{Im}(w)\sim 0$, as $\ell$ grows more QNMs turn to normal modes as shown in Table \ref{tabhighkgamma}.  An interpretation is that the scalar perturbations with high momentum live far away from the horizon in a circular orbit, shielded by the effective potential wall and do not feel the presence of the black string horizon.
\begin{table}
\begin{center}
\begin{tabular}{|c|c|}
\hline
$\ell=10$ & $4.45018,5.55024,6.48684$    \\  \hline

$\ell=20$ & $6.81494,8.21162,9.44172,10.5587,11.5859,12.5322,13.3932,$   \\ \hline
\multicolumn{1}{|c|}{$\ell=50$} & $13.231,15.1091,16.7607,18.2688,19.6739,20.9996,$ \\ 
& $22.2608,23.4676,24.6272,25.7447,26.8238,27.8671,$ \\ 
& $28.8767,29.8534,30.7977,31.709,32.586,33.4256,34.2216,34.9596$ \\ \hline
\end{tabular} 
\caption{The {\it normal-mode} frequencies $w=\omega r_{h}~(r_{h}=3.122)$ of dRGT black string with $M=1,\alpha_g=1,\epsilon=0,\Lambda=-0.01, \gamma=0.4,m=0$.  The momentum parameter $\ell$ is defined as $\ell^{2}\equiv \lambda^{2}+k^{2}/\alpha_{g}^{2}$. }
\label{tabhighkgamma}
\end{center}
\end{table}
\begin{figure}[h]
\centering
 \includegraphics[width=0.7\textwidth]{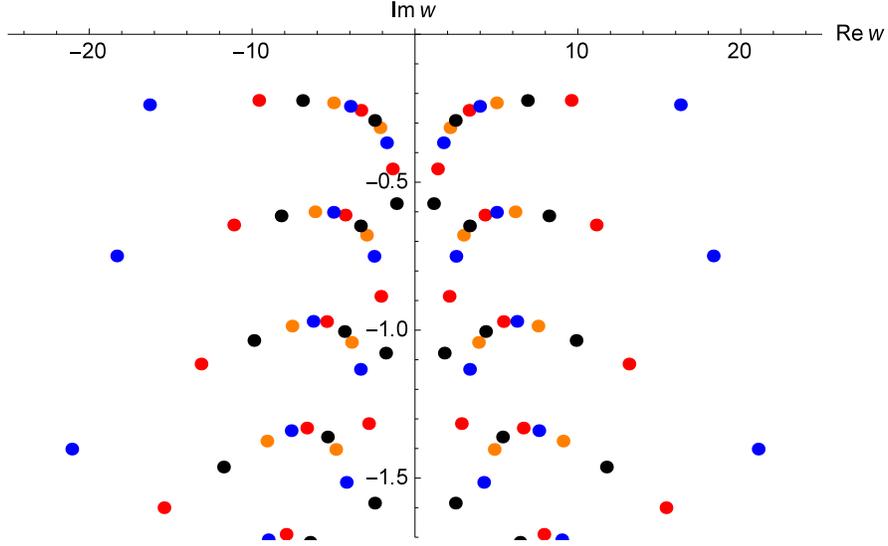}
\caption{The QNMs $w=\omega r_{h}~(r_{h}\in (0.25,5.8))$ of dRGT black string with $M=1,\alpha_g=1,\epsilon=0,\Lambda=-0.01, \lambda =k=0, m=0$ and $\gamma = 0.1,0.2,0.4,0.7,1,2,3,5,10,20,60$~(presented by colored dots from the vertical axis out to both sides) for $N=300$. }
\label{fignegLamb3}
\end{figure}

Interestingly, similar to the black hole in dRGT massive gravity studied in Ref.~\cite{Burikham:2017gdm} when the horizon size is comparable to the characteristic ``AdS'' radius $R_{\Lambda}\equiv\sqrt{1/|\Lambda|}$, the converging QNMs become scarce as shown in Table \ref{tadl} and Fig.~\ref{fignegLamb2}.  On the other hand for black string with small $r_{h}< R_{\Lambda}$, the tower of QNMs appears with excellent convergence.  Figure \ref{fignegLamb3} shows towers of the QNMs of black string for $R_{\Lambda}=10$.  The real and imaginary parts appear to be almost on the straight line~(closer examination shows that it is not exactly a straight line) for each positive value of the massive gravity parameter $\gamma$.  The larger $\gamma$ the more energetic the QNMs become while the lifetime given by the inverse of $\text{Im}(w)$ gets larger at first then smaller for very large $\gamma$.  As explained in Ref.~\cite{Kareeso:2018xum}, the $\gamma r$ term in the spacetime metric represents the constant radial force $-\gamma \hat{r}$ originated from the self-interaction of the massive graviton.  For positive~(negative) $\gamma$, the massive gravity force is attractive~(repulsive).  Naturally for positive $\gamma$, the attractive gravity is stronger so we expect the QNMs to be more energetic, i.e., having larger $\text{Re}(w)$ values.  

In contrast to the $\gamma\geq 0$ case, the lowest QNMs for negative $\gamma$ contain two converging diffusive modes~(i.e. modes with zero $\text{Re}(w)$ and negative $\text{Im}(w)$) as shown in Table \ref{tadsgam}.  Negative $\gamma$ implies antigravity force from massive-graviton self-interaction, the larger the magnitude, the stronger the repulsive force.  Stronger antigravity leads to larger imaginary parts of the diffusive QNMs and thus shorter lifetime.  In constrast to the QNMs of black string in conventional gravity~(see e.g., Ref.\cite{Cardoso:2001vs}), the magnitude of the lowest diffusive QNMs of black string with negative $\gamma$ in massive gravity increases {\it non-linearly} with the horizon~($r_{h}$) and Hawking temperature~($T$) even though $r_{h}$ still increases linearly with $T$, i.e., $r_{h}\sim T\sim \gamma \sim m_{g}^{2}$.  The approximate power-law relation is given by $\text{Im}(w)\sim -|\gamma|^{1.89}$ for the parameters in Table \ref{tadsgam}.
\begin{table}
\begin{center}
\begin{tabular}{|c|c|c|c|c|}
\hline
$\gamma =-0.1$ & $\gamma=-0.2$ & $\gamma=-0.3$ & $\gamma=-0.4$ & $\gamma=-0.5$    \\  \hline

$r_{h}=31.23$ & $r_{h}=60.33$  & $r_{h}=90.15$ & $r_{h}=120.08$ & $r_{h}=150.05$   \\
 
\hline

$T=0.0089$ & $T=0.0162$  & $T=0.0240$ & $T=0.0319$ & $T=0.0398$   \\
 
\hline  
      
$-2.71789i$ & $-9.23955i$  & $-20.4101i$ & $-36.1202i$ & $-56.3462i$   \\
 
\hline
 
$-4.73127i$ & $-16.383i$  & $-36.2566i$  & $-64.1922i$ & $-100.154i$   \\
 
\hline 
\end{tabular} 
\caption{The diffusive QNMs for negative $\gamma, M=1,\Lambda =-0.01, \epsilon = 0, \alpha_{g}=1,m=0,\lambda=k=0$ at $N=300$.  $T$ is the corresponding Hawking temperature of the black string. }
\label{tadsgam}
\end{center}
\end{table}

Holographically, quasinormal frequencies are the poles of the retarded thermal Green function in the dual gauge theory living on the boundary~\cite{Horowitz:1999jd}.  Larger negative imaginary values of QNMs imply that the relaxation time to thermal equilibrium of the perturbed gauge matter in the dual picture is shorter while the temperature is higher due to the antigravity~(since $\gamma$ is negative) effect of massive graviton self-interactions in the bulk.  Scalar perturbation in the bulk is dual to the perturbation of certain scalar operator $\mathcal{O}$ of the gauge theory on the boundary.  The generation of such operator could break scale invariance of the dual field theory.  The perturbed scalar operator will settle to the thermal equilibrium value in the timescale of the relaxation time given by $1/\text{Im}(\omega)$ of the QNMs.  

For holographic massive gravity models~(see e.g., Refs. \cite{Davison:2013jba,Davison:2014lua,Burikham:2016roo,Alberte:2016xja,Hartnoll:2016tri}), the choice of fiducial metric such as (\ref{fideq}) breaks diffeomorphism invariance of the dual gauge theory in the corresponding directions~($z,\varphi$ in our case).  The diffeomorphism-breaking term is effectively the graviton mass term in the gravity theory side.  It generates the momentum dissipation in the broken directions in the dual field theory.  In coherent regime with $\gamma =0$, the {\it sound} poles of QNMs are found as in Table \ref{tadl} and Figure \ref{fignegLamb2}.  Turning on the massive graviton paramater $\gamma>0$ does not generate diffusive modes of QNMs, the sound poles simply move away from the imaginary axis as shown in Fig. \ref{fignegLamb3}.  On the other hand, for $\gamma<0$, the momentum dissipation appears as there are two purely-imaginary modes found, see Table \ref{tadsgam}.  In contrast to the planar geometry case~\cite{Davison:2014lua}, both diffusive poles move down the imaginary axis as $\gamma\sim m_{g}^{2}$ becomes more negative.

\section{Conclusions}\label{sec:conclusion}

In this paper, we have studied the quasinormal modes of massive scalar perturbations on neutral black string background in dRGT massive gravity. The unique characteristics of dRGT black string are the cosmological constant and the linear term $\gamma$ in the metric (\ref{metric}) that are generated naturally from the graviton mass via self-interactions. The $\Lambda$-like term allows one to consider the black string metric separately in either positive and negative $\Lambda$ cases. This point makes the black string in dRGT massive gravity different greatly from the cylindrical black object in general relativity \cite{Lemos:1994xp}. This is because in standard general relativity there is no black string in asymptotically de-Sitter spacetime. We have also examined the horizon structure in each cases. It is found that for positive $\Lambda$, the black string possesses two horizons which we define as event horizon and cosmological horizon. In the negative $\Lambda$ case, the black string only has one event horizon. In a generic set-up where the $\epsilon$ term in the metric (\ref{metric}) is nonzero, the number of horizon structure in negative $\Lambda$ case will be different from ours \cite{Tannukij:2017jtn}.

For the QNMs of a black string with positive $\Lambda$, we start our investigation by considering the spacetime metric (\ref{BSmetric}) in the near extremal limit \cite{Cardoso:2003sw}. In this limit, the cosmological horizon is taken to be very close to the event horizon. It turns out that we can derive the quasinormal frequency analytically by using the P\"{o}schl-Teller technique \cite{Poschl1933}. The frequencies are labeled by the mode number $n$. As $n$ increases, the scalar perturbations decay faster. We find the purely imaginary modes when the scalar mass is turned off. In Sec.\ref{sec:dS}, we perform a fully numerical computation for the QNMs of  black string. The numerical technique called improved asymptotic iteration method is implemented \cite{Cho:2009cj}. Firstly, the results from numerical AIM and analytical formula are compared in the near extremal limit. We find a perfect agreement between the two methods. The results show that the modes with higher oscillation (higher Re$(\omega)$) decay faster than the modes with lower oscillation (lower Re$(\omega)$). In addition, the purely imaginary modes are discovered when the scalar field is massless. However as the scalar mass increases (also the wave number $k$), the quasinormal frequencies appear to be non-diffusive (nonzero real part). We remark that the third-order WKB approximation gives mostly the same results as those obtained from the AIM. We numerically show that the imaginary parts of $\omega$ approach asymptotically to a constant at high momentum $\lambda,k\gg1$. Finally, we find no evidence of any instabilities for all dRGT black string with positive $\Lambda$ since all the perturbation modes investigated in this work decay exponentially.

For negative $\Lambda$ scenario, the black string that is large comparing to the AdS radius $R_{\Lambda}$ has few converging modes at low momentum for $\gamma=0$~(we can choose $M$ to be small to get small black string though).  As the momentum of the scalar field gets higher, either with $k$ or $\lambda$, more converging modes appear at exactly the same $\text{Re}(w)$ with larger negative $\text{Im}(w)$.  For large momentum limit, we numerically establish asymptotic relation between real and imaginary parts of the QNMs consistent with the WKB method found in Ref.~\cite{Festuccia:2008zx}.  When the massive graviton effect $\gamma\sim m_{g}^{2}$ is turned on to positive value, the QNMs becomes sequences of normal modes for high momenta.  The higher the momentum, the more number of normal modes appear as shown in Table \ref{tabhighkgamma}.  At zero momentum for a fixed mass, positive $\gamma$  makes black-string horizon smaller and there is a number of QNMs found.  As the effect of massive graviton gets stronger, the energy of the QNMs grows larger, the lifetime becomes longer~(i.e., $\text{Im}(w)$ becomes less negative) at first then starts to get shorter for very large $\gamma$.

When $\Lambda<0, \gamma<0$, the massive graviton self-interactions generate antigravity at long distances.  With such antigravity, the QNMs remarkably become diffusive with $\text{Re}(w)=0$.  The relaxation time is shorter with stronger antigravity.  Holographically, the choice of fiducial metric (\ref{fideq}) breaks diffeomorphism invariance in the $z,\varphi$ directions of the perturbations resulting in momentum dissipation in the dual field theory.  We note the importance of sign of $\gamma$ in the movement of sound poles to collide and form diffusive poles, as $\gamma$ changes from positive to zero and to negative values.  Long-distance gravity changes from attractive to repulsive as the dual hydrodynamics changes from coherent to momentum-dissipative regime. 

As a generalization of this work, one can investigate the effect of linear charged scalar perturbation on charged dRGT black string spacetime.  With the existence of black string and scalar charges, this could possibly introduce the effect of superradiance \cite{Brito:2015oca} on this type of spacetime background.  Holographically, unstable charged scalar perturbations in such charged-string~(asymptotically AdS) background signals the superconducting phase transition of the dual field system on the cylindrical boundary that resembles e.g., the carbon nanotube.

\acknowledgments

We would like to thank the anonymous referees for valuable comments and suggestions that helped us improve our manuscript.  S.P. is supported by Rachadapisek Sompote Fund for Postdoctoral Fellewship, Chulalongkorn University.  P.B. is supported in part by the Thailand Research Fund~(TRF), Office of Higher Education Commission (OHEC) and Chulalongkorn University under grant RSA6180002.  L.T. is supported by the National Research Foundation of Korea (NRF) grant funded by the Korea government (MSIP) (No.2016R1C1B1010107). 


\begin{thebibliography}{99}

\bibitem{Fierz:1939ix} 
  M.~Fierz and W.~Pauli,
  \emph{``On relativistic wave equations for particles of arbitrary spin in an electromagnetic field,''}
  Proc.\ Roy.\ Soc.\ Lond.\ A {\bf 173}, 211 (1939).

\bibitem{Hinterbichler:2011tt} 
  K.~Hinterbichler,
  \emph{``Theoretical Aspects of Massive Gravity,''}
  Rev.\ Mod.\ Phys.\  {\bf 84}, 671 (2012)
  [arXiv:1105.3735 [hep-th]].

\bibitem{vanDam:1970vg} 
  H.~van Dam and M.~J.~G.~Veltman,
  \emph{``Massive and massless Yang-Mills and gravitational fields,''}
  Nucl.\ Phys.\ B {\bf 22}, 397 (1970).

\bibitem{Zakharov:1970cc} 
  V.~I.~Zakharov,
  \emph{``Linearized gravitation theory and the graviton mass,''}
  JETP Lett.\  {\bf 12}, 312 (1970)
  [Pisma Zh.\ Eksp.\ Teor.\ Fiz.\  {\bf 12}, 447 (1970)].

\bibitem{Vainshtein:1972sx} 
  A.~I.~Vainshtein,
  \emph{``To the problem of nonvanishing gravitation mass,''}
  Phys.\ Lett.\  {\bf 39B}, 393 (1972).

\bibitem{Boulware:1973my} 
  D.~G.~Boulware and S.~Deser,
  \emph{``Can gravitation have a finite range?,''}
  Phys.\ Rev.\ D {\bf 6}, 3368 (1972).

\bibitem{deRham:2010ik} 
  C.~de Rham and G.~Gabadadze,
  \emph{``Generalization of the Fierz-Pauli Action,''}
  Phys.\ Rev.\ D {\bf 82}, 044020 (2010)
  [arXiv:1007.0443 [hep-th]].
  
\bibitem{deRham:2010kj} 
  C.~de Rham, G.~Gabadadze and A.~J.~Tolley,
  \emph{``Resummation of Massive Gravity,''}
  Phys.\ Rev.\ Lett.\  {\bf 106}, 231101 (2011)
  [arXiv:1011.1232 [hep-th]].

\bibitem{Gumrukcuoglu:2011ew} 
  A.~E.~Gumrukcuoglu, C.~Lin and S.~Mukohyama,
  \emph{``Open FRW universes and self-acceleration from nonlinear massive gravity,''}
  JCAP {\bf 1111}, 030 (2011)
  [arXiv:1109.3845 [hep-th]].

\bibitem{Gumrukcuoglu:2011zh} 
  A.~E.~Gumrukcuoglu, C.~Lin and S.~Mukohyama,
  \emph{``Cosmological perturbations of self-accelerating universe in nonlinear massive gravity,''}
  JCAP {\bf 1203}, 006 (2012)
  [arXiv:1111.4107 [hep-th]].
\bibitem{Ghosh:2015cva} 
  S.~G.~Ghosh, L.~Tannukij and P.~Wongjun,
  \emph{``A class of black holes in dRGT massive gravity and their thermodynamical properties,''}
  Eur.\ Phys.\ J.\ C {\bf 76}, no. 3, 119 (2016)
  [arXiv:1506.07119 [gr-qc]].

\bibitem{Kodama:2013rea} 
  H.~Kodama and I.~Arraut,
  PTEP {\bf 2014}, 023E02 (2014)
  doi:10.1093/ptep/ptu016
  [arXiv:1312.0370 [hep-th]].

\bibitem{Babichev:2013una} 
  E.~Babichev and A.~Fabbri,
  Class.\ Quant.\ Grav.\  {\bf 30}, 152001 (2013)
  doi:10.1088/0264-9381/30/15/152001
  [arXiv:1304.5992 [gr-qc]].
  
\bibitem{Babichev:2014oua} 
  E.~Babichev and A.~Fabbri,
  Phys.\ Rev.\ D {\bf 89}, no. 8, 081502 (2014)
  doi:10.1103/PhysRevD.89.081502
  [arXiv:1401.6871 [gr-qc]].  
   
\bibitem{Lemos:1994xp}
  J.~P.~S.~Lemos,
  \emph{``Cylindrical black hole in general relativity,''}
  Phys.\ Lett.\ B {\bf 353} (1995) 46
  [gr-qc/9404041].

\bibitem{Konoplya:2008yy} 
  R.~A.~Konoplya, K.~Murata, J.~Soda and A.~Zhidenko,
  Phys.\ Rev.\ D {\bf 78}, 084012 (2008)
   [arXiv:0807.1897 [hep-th]].  

\bibitem{Tannukij:2017jtn}
  L.~Tannukij, P.~Wongjun and S.~G.~Ghosh,
  \emph{``Black String in dRGT Massive Gravity,''}
  Eur.\ Phys.\ J.\ C {\bf 77} (2017) no.12,  846
  [arXiv:1701.05332 [gr-qc]].
  
  
\bibitem{Prasia:2016fcc}
  P.~Prasia and V.~C.~Kuriakose,
  \emph{``Quasi Normal Modes and P-V Criticallity for scalar perturbations in a class of dRGT massive gravity around Black Holes,''}
  Gen.\ Rel.\ Grav.\  {\bf 48} (2016) no.7,  89
  [arXiv:1606.01132 [gr-qc]].
  
\bibitem{Burikham:2017gdm}
  P.~Burikham, S.~Ponglertsakul and L.~Tannukij,
  \emph{``Charged scalar perturbations on charged black holes in de Rham-Gabadadze-Tolley massive gravity,''}
  Phys.\ Rev.\ D {\bf 96} (2017) no.12,  124001
  [arXiv:1709.02716 [gr-qc]].

\bibitem{Kokkotas:1999bd}
  K.~D.~Kokkotas and B.~G.~Schmidt,
  \emph{``Quasinormal modes of stars and black holes,''}
  Living Rev.\ Rel.\  {\bf 2} (1999) 2
  [gr-qc/9909058].

\bibitem{Berti:2009kk}
  E.~Berti, V.~Cardoso and A.~O.~Starinets,
  \emph{``Quasinormal modes of black holes and black branes,''}
  Class.\ Quant.\ Grav.\  {\bf 26} (2009) 163001
  [arXiv:0905.2975 [gr-qc]].

\bibitem{Konoplya:2011qq}
  R.~A.~Konoplya and A.~Zhidenko,
  \emph{``Quasinormal modes of black holes: From astrophysics to string theory,''}
  Rev.\ Mod.\ Phys.\  {\bf 83} (2011) 793
  [arXiv:1102.4014 [gr-qc]].

\bibitem{Davison:2013jba} 
  R.~A.~Davison,
  \emph{``Momentum relaxation in holographic massive gravity,''}
  Phys.\ Rev.\ D {\bf 88}, 086003 (2013)
  [arXiv:1306.5792 [hep-th]].
  
\bibitem{Davison:2014lua} 
  R.~A.~Davison and B.~Goutéraux,
  \emph{``Momentum dissipation and effective theories of coherent and incoherent transport,''}
  JHEP {\bf 1501}, 039 (2015)
  [arXiv:1411.1062 [hep-th]].  
   
\bibitem{Cardoso:2001vs}
  V.~Cardoso and J.~P.~S.~Lemos,
  \emph{``Quasinormal modes of toroidal, cylindrical and planar black holes in anti-de Sitter space-times,''}
  Class.\ Quant.\ Grav.\  {\bf 18} (2001) 5257
  [gr-qc/0107098].
  
\bibitem{1stfootnote}
The black string could posses three positive real roots when $\epsilon$ is non-vanishing, see Ref.~\cite{Tannukij:2017jtn}.

\bibitem{AIM:2003}
	H.~Ciftci, R.~L.~Hall and N.~Saad,
	\emph{``Asymptotic iteration method for eigenvalue problems,''}
	J. Phys. {\bf A 36} (2003) 11807.
	
\bibitem{Cho:2009cj} 
  H.~T.~Cho, A.~S.~Cornell, J.~Doukas and W.~Naylor,
  \emph{``Black hole quasinormal modes using the asymptotic iteration method''},
  Class.\ Quant.\ Grav.\  {\bf 27}, 155004 (2010)
  [arXiv:0912.2740 [gr-qc]].

\bibitem{Konoplya:2003ii} 
  R.~A.~Konoplya,
  Phys.\ Rev.\ D {\bf 68}, 024018 (2003)
  [gr-qc/0303052].

\bibitem{Matyjasek:2017psv} 
  J.~Matyjasek and M.~Opala,
  Phys.\ Rev.\ D {\bf 96}, no. 2, 024011 (2017)
  doi:10.1103/PhysRevD.96.024011
  [arXiv:1704.00361 [gr-qc]].
    
\bibitem{Zangeneh:2017rhc} 
  M.~Kord Zangeneh, B.~Wang, A.~Sheykhi and Z.~Y.~Tang,
  \emph{``Charged scalar quasi-normal modes for linearly charged dilaton-Lifshitz solutions,''}
  Phys.\ Lett.\ B {\bf 771}, 257 (2017)
  [arXiv:1701.03644 [hep-th]]. 

\bibitem{Bonazzola:1998qx} 
  S.~Bonazzola, E.~Gourgoulhon and J.~A.~Marck,
  Phys.\ Rev.\ D {\bf 58}, 104020 (1998)
  doi:10.1103/PhysRevD.58.104020
  [astro-ph/9803086].
  
\bibitem{Ferrari:1984zz}
  V.~Ferrari and B.~Mashhoon,
  \emph{``New approach to the quasinormal modes of a black hole,''}
  Phys.\ Rev.\ D {\bf 30} (1984) 295.
  
\bibitem{Cardoso:2003sw}
  V.~Cardoso and J.~P.~S.~Lemos,
  \emph{``Quasinormal modes of the near extremal Schwarzschild-de Sitter black hole,''}
  Phys.\ Rev.\ D {\bf 67} (2003) 084020
  [gr-qc/0301078].

\bibitem{Konoplya:2004wg} 
  R.~A.~Konoplya and A.~V.~Zhidenko,
  Phys.\ Lett.\ B {\bf 609}, 377 (2005)
  [gr-qc/0411059].

\bibitem{Konoplya:2006br} 
  R.~A.~Konoplya and A.~Zhidenko,
  Phys.\ Rev.\ D {\bf 73}, 124040 (2006)
  [gr-qc/0605013].
  
\bibitem{Poschl1933}
	G.~"P{\"o}schl and E.~Teller,
	\emph{``Bemerkungen zur Quantenmechanik des anharmonischen Oszillators,"}
	Zeitschrift f{\"u}r Physik (1993)

\bibitem{Zhidenko:2003wq} 
  A.~Zhidenko,
  \emph{``Quasinormal modes of Schwarzschild de Sitter black holes,''}
  Class.\ Quant.\ Grav.\  {\bf 21}, 273 (2004)
  [gr-qc/0307012].
  
\bibitem{Molina:2003ff}
  C.~Molina,
  \emph{``Quasinormal modes of d-dimensional spherical black holes with near extreme cosmological constant,''}
  Phys.\ Rev.\ D {\bf 68} (2003) 064007
  [gr-qc/0304053].

\bibitem{Fernando:2014gda}
  S.~Fernando and T.~Clark,
  \emph{``Black holes in massive gravity: quasi-normal modes of scalar perturbations,''}
  Gen.\ Rel.\ Grav.\  {\bf 46} (2014) no.12,  1834
  [arXiv:1411.6537 [gr-qc]].
  
\bibitem{2ndfootnote}
Setting $\gamma=0,\epsilon=1,\alpha_g=2,m_s=0$ and let $\displaystyle{\lambda^2+\frac{k^2}{4}=l(l+1)}$. 
 
\bibitem{Moss:2001ga} 
  I.~G.~Moss and J.~P.~Norman,
  \emph{``Gravitational quasinormal modes for anti-de Sitter black holes''},
  Class.\ Quant.\ Grav.\  {\bf 19}, 2323 (2002)
  [arXiv:0201016 [gr-qc]].

\bibitem{WadeNB} 
  W.~Naylor,
  \emph{``Asymptotic Iteration Method for Black Hole Physics''},
  http://www-het.phys.sci.osaka-u.ac.jp/~naylor/AIM.html
  [Online; accessed 18-February-2018].
   
\bibitem{Schutz:1985zz} 
  B.~F.~Schutz and C.~M.~Will,
  \emph{``Black Hole Normal Modes: A Semianalytic Approach,''}
  Astrophys.\ J.\  {\bf 291}, L33 (1985).

\bibitem{Iyer:1986np} 
  S.~Iyer and C.~M.~Will,
  \emph{``Black Hole Normal Modes: A {WKB} Approach. 1. Foundations and Application of a Higher Order {WKB} Analysis of Potential Barrier Scattering,''}
  Phys.\ Rev.\ D {\bf 35}, 3621 (1987).

\bibitem{Cardoso:2008bp} 
  V.~Cardoso, A.~S.~Miranda, E.~Berti, H.~Witek and V.~T.~Zanchin,
  Phys.\ Rev.\ D {\bf 79}, 064016 (2009)
  doi:10.1103/PhysRevD.79.064016
  [arXiv:0812.1806 [hep-th]].

\bibitem{yaffec}
Sample codes can be found in http://msstp.org/?q=node/289.

\bibitem{Festuccia:2008zx} 
  G.~Festuccia and H.~Liu,
  \emph{``A Bohr-Sommerfeld quantization formula for quasinormal frequencies of AdS black holes,''}
  Adv.\ Sci.\ Lett.\  {\bf 2}, 221 (2009)
  [arXiv:0811.1033 [gr-qc]].  

\bibitem{Kareeso:2018xum}
  P.~Kareeso, P.~Burikham and T.~Harko,
  \emph{``Mass - radius ratio bounds for compact objects in Massive Gravity theory,''}
  arXiv:1802.01017 [gr-qc].

\bibitem{Horowitz:1999jd} 
  G.~T.~Horowitz and V.~E.~Hubeny,
  \emph{``Quasinormal modes of AdS black holes and the approach to thermal equilibrium,''}
  Phys.\ Rev.\ D {\bf 62}, 024027 (2000)
  [hep-th/9909056].
  
\bibitem{Burikham:2016roo} 
  P.~Burikham and N.~Poovuttikul,
  \emph{``Shear viscosity in holography and effective theory of transport without translational symmetry,''}
  Phys.\ Rev.\ D {\bf 94}, no. 10, 106001 (2016)
  [arXiv:1601.04624 [hep-th]].
    
\bibitem{Alberte:2016xja} 
  L.~Alberte, M.~Baggioli and O.~Pujolas,
  \emph{``Viscosity bound violation in holographic solids and the viscoelastic response,''}
  JHEP {\bf 1607}, 074 (2016)
  [arXiv:1601.03384 [hep-th]].
      
\bibitem{Hartnoll:2016tri} 
  S.~A.~Hartnoll, D.~M.~Ramirez and J.~E.~Santos,
  \emph{``Entropy production, viscosity bounds and bumpy black holes,''}
  JHEP {\bf 1603}, 170 (2016)
  [arXiv:1601.02757 [hep-th]].
  
\bibitem{Brito:2015oca} 
  R.~Brito, V.~Cardoso and P.~Pani,
  \emph{``Superradiance : energy extraction, black-hole bombs and implications for astrophysics and particle physics''},
  Lect.\ Notes Phys.\  {\bf 906}, pp.1 (2015)
  [arXiv:1501.06570 [gr-qc]].
  






                          
            
\end{thebibliography}

\end{document}